\documentclass[a4paper]{article}

\usepackage[pages=all, color=black, position={current page.south}, placement=bottom, scale=1, opacity=1, vshift=5mm]{background}
\backgroundsetup{contents={}}

\usepackage{algorithm2e}

\usepackage{multirow}
\usepackage{subcaption}
\usepackage{comment}
\usepackage{todonotes}

\usepackage{natbib,amsmath,amssymb}
\usepackage[margin=1.2in]{geometry} 

\usepackage{amsmath}
\newcommand{\argmin}{\ensuremath{\operatorname{argmin}}}

\title{Transonic Buffet Modeling via Invariant Manifolds}

\author{Tea Vojkovi\'c\textsuperscript{1,2}, David Quero\textsuperscript{3},  Rahul Jayaraj\textsuperscript{1,3},
Christoph Kaiser\textsuperscript{3}, \\
Dimitris Boskos\textsuperscript{2}, Abel-John Buchner\textsuperscript{1}\\ \\
\small \textsuperscript{1}Laboratory for Aero and Hydrodynamics,\\
\small Delft University of Technology, Delft, The Netherlands\\
\small \textsuperscript{2}Delft Center for Systems and Control,\\
\small Delft University of Technology, Delft, The Netherlands\\
\small \textsuperscript{3}DLR (German Aerospace Center), Institute of Aeroelasticity, G\"{o}ttingen, Germany\\ \\
\small Corresponding author: a.j.buchner@tudelft.nl}

\date{}

\begin{document}
\maketitle

\begin{abstract}

In transonic flow over aircraft wings, shock–boundary-layer interactions can give rise to transonic buffet, which degrades maneuverability through unsteady aerodynamic loads. 
Beyond its practical importance, two-dimensional transonic buffet represents a canonical example of a global instability for which reduced-order modeling remains challenging due to nonlinearity, sharp spatial gradients, and the coexistence of an unstable equilibrium with an attracting limit cycle. Commonly, reduced-order models of such phenomena capture nonlinear dynamics only in aerodynamic observables, while prediction of the full flow state is achieved through linear representations valid only near the unstable equilibrium or on the limit cycle.

In this work, we present a reduced-order model that predicts the nonlinear evolution of the full flow field by exploiting the existence of an attracting two-dimensional invariant manifold. We adapt an existing data-driven framework for identifying invariant manifolds and the associated reduced dynamics, making it suitable for scaling to large-scale CFD applications. The invariant manifold is identified as a graph over its tangent space using an iterative encoder-update and the reduced dynamics are obtained via least-squares regression. A subsequent extended normal-form transformation enables physical interpretability of the model through a modal decomposition of the flow.

The reduced-order model is identified for transonic buffet over the OAT15A supercritical airfoil, showing that it is possible to achieve this accurately using just a single training trajectory. Validation against independent simulations demonstrates accurate prediction of nonlinear behavior, together with reliable reconstruction of the full flow field, particularly in the late-transient and limit-cycle regimes.

\end{abstract}

\section{Introduction}

In transonic flow over aircraft wings, shock-boundary-layer interactions can generate a self-sustained unsteadiness known as transonic buffet. Buffet degrades maneuverability through unsteady aerodynamic loads and may trigger structural limit-cycle oscillations that impose damaging cyclic stresses. Beyond its practical importance, buffet is a canonical example of a nonlinear global instability whose prediction and control remain challenging, as reviewed comprehensively by~\cite{giannelis2017review}.

Buffeting flows exhibit two global instabilities: a two-dimensional (2D) streamwise mode identified by~\cite{crouch2009origin} and a three-dimensional (3D) spanwise mode reported by~\cite{iovnovich2015numerical},~\cite{paladini2019transonic}, and~\cite{lusher2025implicit}. The 2D mode causes low-frequency streamwise shock oscillations,  
while the 3D mode occurs only on fully three-dimensional wings where spanwise extent or sweep enables cross-flow unsteadiness. The two modes can coexist, showing that 2D mechanisms remain fundamental even in three-dimensional buffet (\cite{paladini2019transonic}).

Two-dimensional airfoils are therefore widely used as benchmark configurations for studying transonic buffet. The literature distinguishes between Type I buffet, involving shock oscillations on both sides of an symmetric airfoil at zero angle of attack, and Type II buffet, which arises at non-zero incidence on supercritical asymmetric airfoils.
Depending on the boundary-layer state upstream of the shock, it may further manifest as laminar buffet~(\cite{dandois2018large,zauner2020wide,moise2022large,song2024numerical}) or turbulent buffet~(\cite{fukushima2018wall,cuong2022large}).  
Motivated by the relevance of high-Reynolds-number flows over commercial wings, we focus here on Type~II turbulent buffet. In this regime, transonic buffet occurs only within a narrow range of Mach numbers and angles of attack and is dominated by a self-sustained low-frequency oscillation (\cite{lee1990oscillatory,raveh2009numerical,giannelis2018influence,accorinti2022experimental,nitzsche2022effect}). Global stability analyses show that, for a fixed Mach number, buffet onset occurs through a Hopf bifurcation of the steady base flow as the angle of attack exceeds a critical value~(\cite{crouch2009origin,sartor2015stability,crouch2024weakly}). This two-dimensional global instability leads to a periodic limit cycle characterized by coherent shock oscillations on the suction side of the airfoil, synchronized with the thickening and thinning of the downstream separated boundary layer. The resulting low-frequency oscillations have been consistently reproduced in numerical simulations (\cite{raveh2009numerical,crouch2009origin,fukushima2018wall,paladini2019transonic,eldridge2024comparison}) and observed in wind-tunnel experiments (\cite{mcdevitt1985static,d2021experimental,schauerte2023experimental}).
Medium-frequency von Kármán and Kelvin--Helmholtz instabilities may also develop within the separated region, but their energy is more than two orders of magnitude lower than that of the dominant global mode, as shown by~\cite{giannelis2017review}, confirming the global mode's dominant role in the flow dynamics.

Transonic buffet poses major challenges for prediction, modeling, and control due to its nonlinearity, sharp spatial gradients associated with shock–boundary-layer interactions, and the coexistence of an unstable equilibrium with an attracting limit cycle. At turbulent Reynolds numbers, numerical simulation of the compressible Navier--Stokes equations yields state spaces of $\mathcal{O}(10^6)$ degrees of freedom or more, making direct control design computationally prohibitive. Reduced-order modeling is therefore essential to enable tractable analysis, prediction, and control, while retaining the key physics of buffet dynamics.

A wide range of reduced-order modeling approaches has been proposed to describe buffet unsteadiness, differing in whether they can reconstruct the full flow field or only capture aerodynamic observables, in their treatment of nonlinear dynamics, and in the physical interpretability of the resulting reduced models.
 
Reduced-order models (ROMs) based solely on aerodynamic observables aim to predict integrated quantities such as lift and moment coefficients without reconstructing the full flow field. Examples include the linear auto-regressive (ARX) models of~\cite{gao2016new,gao2017mechanism} and the weakly nonlinear framework of~\cite{crouch2024weakly}. While the ARX models considered in~\cite{gao2016new,gao2017mechanism} can predict linear transient input–output behavior near the unstable equilibrium, they do not capture nonlinear saturation leading to a limit cycle and lack physical interpretability. In contrast, the weakly nonlinear approach of~\cite{crouch2024weakly}, formulated via a Stuart--Landau equation, captures the essential nonlinear dynamics and provides physical insight grounded in bifurcation theory, but remains restricted to scalar observables. 

Approaches that aim to reconstruct the full flow state generally offer greater physical insight. Proper Orthogonal Decomposition (POD;~\cite{lumley1967structure}) provides an efficient low-dimensional representation of flow fields and has been applied to transonic buffet by~\cite{poplingher2019modal} and~\cite{iwatani2022pod}, but is not itself a dynamical model and cannot predict temporal evolution. Moreover, flows with sharp gradients such as shocks typically require a large number of POD modes for accurate reconstruction (\cite{li2016performance}). A recent autoencoder-based framework proposed in~\cite{fukami2025compact} demonstrated improved reconstruction accuracy compared to POD, but remains limited to static modeling. To incorporate dynamics, Dynamic Mode Decomposition (DMD; \cite{rowley2009spectral}) provides a linear approximation of flow evolution and has been applied near the unstable base flow (\cite{gao2017active}) and to approximate the saturated limit cycle (\cite{poplingher2019modal,feldhusen2021analysis,das2023transonic,weiner2023robust}). Linearization about the base flow is restricted to early linear transients in the vicinity of the unstable equilibrium, while treating the limit cycle as a solution of a linear oscillatory system leads to strong sensitivity to initial conditions.
 Neither approach captures the nonlinear settlement onto the limit cycle.  
A hybrid formulation combining a nonlinear observable model based on a Stuart--Landau equation with DMD was introduced by \cite{sansica2022system}, enabling full-state reconstruction in the periodic regime but not during transient evolution.

To the best of our knowledge, no existing reduced-order model can both predict nonlinear buffet dynamics and reconstruct the full flow state across the phase space from the unstable equilibrium to the saturated limit cycle. This gap motivates us to explore a modeling approach based on invariant-manifold theory. For systems undergoing a Hopf bifurcation, the long-term dynamics are confined to a two-dimensional, attracting invariant manifold attached to the equilibrium (\cite{kuznetsov1998elements}) and can be described by reduced dynamics evolving on it. The invariant manifold further defines a mapping from reduced to the full state space, acting as a decoder that enables reconstruction of the complete flow field. Thus this approach provides a promising reduced-order modeling framework for transonic buffet.

A powerful framework for establishing the existence and regularity of invariant manifolds associated with fixed points of dynamical systems is the parameterization method, originally introduced in \cite{cabre2003parameterization-a,cabre2003parameterization-b}. A comprehensive exposition of the theory, along with numerous computational aspects, is provided in the monograph \cite{haro2016parameterization}. The parameterization method guarantees the existence of invariant manifolds under general spectral assumptions on the linearization at the equilibrium. More recently, it has been leveraged as a model reduction tool in a series of works (\cite{haller2016nonlinear,jain2022compute,cenedese2022data}), which further relax earlier stability restrictions imposed on the spectral subspaces of the linearization (\cite{haller2023nonlinear}).

Both the invariant manifold and its internal dynamics can be determined directly from the governing equations. This approach has been successfully applied to fluid flows to obtain a reduced-order model of the cylinder wake (\cite{carini2015centre}). Several data-driven alternatives have also been introduced, including approaches based on spectral submanifolds (SSMs) (\cite{cenedese2022data}) and methods relying on invariant foliations (\cite{szalai2020invariant,szalai2023data}), which allow identification of both the manifold and its reduced dynamics directly from simulation or experimental data. Although SSM-based techniques have demonstrated success for moderate-dimensional mechanical systems and incompressible flows of $\mathcal{O}(10^{5})$ degrees of freedom including cylinder wake (\cite{cenedese2022data}), pipe flow~ (\cite{kaszas2024capturing}) and sloshing-tank dynamics~ (\cite{axaas2023fast}), their application to unsteady transonic flows, and to transonic buffet in particular, remains largely unexplored.

Applying data-driven invariant-manifold-based frameworks to transonic buffet modeling entails several practical challenges addressed in this work. These stem from (i) the extreme dimensionality of the governing system and (ii) the impracticality of obtaining data across the entire relevant region of phase space. While the first point has been discussed above, the second arises because capturing the nonlinear evolution from the unstable equilibrium to the saturated limit cycle requires either very long transients in a single simulation or multiple independently initialised simulations, both of which are prohibitively expensive. Since accurate modeling of the transient and limit-cycle dynamics already demands substantial data in these regions, it is generally impractical to also obtain data sufficiently close to the unstable equilibrium, motivating methodologies that do not rely on such information.

A central challenge in data-driven invariant-manifold identification is the definition of suitable reduced coordinates, typically achieved through an encoder that maps the full state onto a reduced space. Existing approaches either identify linear encoders and decoders simultaneously by solving a nonconvex autoencoding problem~(\cite{cenedese2022data}), or employ nonlinear encoders associated with invariant foliations~(\cite{szalai2020invariant,szalai2023data}), both of which become computationally expensive in high-dimensional settings. A substantial simplification arises when a suitable linear encoder or reduced coordinates are available \emph{a priori}, as this reduces manifold identification to a convex least-squares problem. This is the case for the cylinder wake, where the most energetic POD modes on the limit cycle are known to form a suitable reduced basis (\cite{noack2003hierarchy}), and were subsequently employed as a linear encoder for invariant-manifold identification in~\cite{cenedese2022data}.
 In practice, however, learning a suitable linear encoder typically relies on data sufficiently close to the equilibrium (\cite{axaas2023fast,bettini2025data,buurmeijer2025taming}), while prior knowledge of appropriate reduced coordinates is generally unavailable for high-dimensional transonic buffet flows.

To address these challenges, we adapt the invariant-manifold-based approach of~\cite{cenedese2022data} by modifying the manifold-identification procedure. The proposed approach identifies the invariant manifold as a graph over its tangent space by iteratively updating  a linear encoder. Starting from an initial encoder, the manifold is first identified via least-squares regression, after which the encoder is updated using the linear part of the resulting parameterization. Repeating this procedure yields convergence to an orthogonal graph-style parameterization. Each iteration involves only linear regression whose size depends on the number of snapshots and the polynomial degree of the representation, and is independent of the full state dimension. The approach is therefore suitable for high-dimensional CFD data and compatible with limited or unavailable samples near the steady base flow, without requiring prior knowledge of suitable reduced coordinates. Once the manifold is identified, the reduced dynamics on it are identified from projected trajectory data via least-squares regression.

The approach is further complemented by an extended normal-form transformation, introduced in~\cite{cenedese2022data}, which yields a simplified representation of the reduced dynamics. For systems undergoing a Hopf bifurcation, the reduced dynamics can be expressed in a simple amplitude–phase form, when truncated at third order, corresponding to the classical Stuart--Landau equation, using this transformation. We show that, for such systems, rewriting the manifold parameterization in normal-form coordinates enables a physically interpretable modal decomposition in which the full dynamics is described by modes oscillating at distinct characteristic frequencies. This representation remains valid throughout the transient evolution from the unstable equilibrium to the saturated limit cycle, under the assumption of slow amplitude evolution. Using this approach, we identify a two-dimensional reduced-order model of transonic buffet that predicts the nonlinear evolution from the unstable equilibrium to the saturated limit cycle, reconstructs the full flow field, and provides a physically interpretable modal decomposition.

The paper is organized as follows. Section~\ref{sec:problem} introduces the governing equations describing the flow, highlights the key dynamical features relevant to reduced-order modeling, and outlines the objectives of this work.
Section~\ref{sec:ROM} presents the theoretical background on invariant manifolds and details the proposed data-driven reduced-order modeling approach. Section~\ref{sec:Results} presents the reduced-order model constructed for transonic buffet over the OAT15A supercritical airfoil. Finally, Section~\ref{sec:Conclusion} summarizes the main findings and outlines directions for future work.

\section{Problem Description}\label{sec:problem}

We consider two-dimensional transonic flow over an airfoil exhibiting turbulent shock buffet. In the following, we introduce the governing equations used to simulate this flow and highlight the key dynamical characteristics that will later be exploited for reduced-order modeling.

\subsection{Governing Equations}

Under the high-Reynolds-number conditions relevant to practical aeronautical applications, the boundary layer is fully turbulent, and the large-scale unsteadiness associated with buffet can be modeled using the unsteady Reynolds-averaged Navier--Stokes (URANS) equations closed with an appropriate turbulence model. Reynolds averaging is justified because the characteristic timescale of the low-frequency shock oscillation is much longer than that of the small-scale turbulent fluctuations in the boundary layer (\cite{sartor2015stability}). We note, however, that any potential medium-frequency instabilities developing within the separated boundary layer (\cite{giannelis2017review}) are unlikely to be captured by URANS (\cite{poplingher2019modal}). 

The two-dimensional URANS equations can be written as a system of partial differential equations (PDEs)
\begin{equation}\label{eq:URANS}
  \frac{\partial q(\mathbf{x},t)} {\partial t} = \mathcal{F}(q, Re, M_{\infty}, \alpha)
\end{equation}
where $\mathbf{x} \in \Omega \in \mathbb{R}^2$ denotes the spatial coordinate and $\mathcal{F}$ denotes the nonlinear residual operator incorporating all fluxes, assumed to be twice continuously differentiable over the domain of interest. \color{black}
The state vector $ q=[\rho, \rho \mathbf{u}, \rho E, \mathbf{q}_{\text{tf}}]$ consists of Reynolds-averaged conservative variables, namely, the density $\rho$, the momentum $\rho \mathbf{u}$, where  $\mathbf{u}(\mathbf{x},t)=[u, v]$ is the velocity, the total energy $\rho E$, and the turbulence-related variables $\mathbf{q}_{\text{tf}}$. 
Using the one-equation Spalart–Allmaras model (\cite{spalart1992one}),
$\mathbf{q}_{\text{tf}} = \rho \tilde{\nu}$,
with $\tilde{\nu}$ being the modified turbulent kinematic viscosity. 
The equations~\eqref{eq:URANS} depend on different parameters including Reynolds number $Re$, the angle of attack $\alpha$, and the freestream Mach number $M_{\infty}$. The steady solution $q_b$ of~\eqref{eq:URANS} satisfying
\begin{equation*}
    \mathcal{F}(q_b, Re, M_{\infty}, \alpha)=\mathbf{0},
\end{equation*}
is referred to as the \textit{base flow}. The base flow is also parameter-dependent. 
Since the governing equations include a turbulence model, the base flow incorporates the averaged influence of high-frequency turbulent fluctuations through the modeled closure terms.
Considering an unsteady perturbation $q'=q-q_b$ of the base flow, the governing equations become
\begin{equation}\label{eq:URANSunsteady}
  \frac{\partial q'}{\partial t} = \mathcal{F}(q'+q_b, Re, M_{\infty}, \alpha). 
\end{equation}
By construction, $q'=0$ corresponds to a fixed point of this system, representing the base flow $q_b$. The unsteady state $q'$ therefore represents the low-frequency, large-scale variations of the flow.

\subsection{Flow Dynamics}

After spatial discretization, the semi-discrete URANS equations yield a system of ordinary differential equations (ODEs)
\begin{equation}\label{eq:URANSODE}
  \frac{d \mathbf{q}}{d t} = \mathbf f(\mathbf q)\equiv\mathbf{f}(\mathbf{q};\mathbf{q}_b, \mathbf{p}).
\end{equation} 
Here, $\mathbf{q}(t)\in\mathbb{R}^n$ denotes the $n$-dimensional discretized unsteady state and
$\mathbf{q}_b\in\mathbb{R}^n$ the corresponding steady base flow state. The nonlinear vector field
$\mathbf{f}:\mathbb{R}^n\rightarrow\mathbb{R}^n$ incorporates all discretized fluxes and boundary
conditions arising from $\mathcal{F}(\mathbf{q}+\mathbf{q}_b,\mathbf{p})$.
By construction, $\mathbf{q}=\mathbf{0}$ corresponds to the steady base flow solution, i.e. a fixed point of the dynamical system, around which the unsteady perturbation dynamics evolve.
Equation~\eqref{eq:URANSODE} therefore represents a high-dimensional nonlinear dynamical system whose parameters $\mathbf{p}=[M_{\infty}, \alpha]\in \mathbb{R}^2$ govern the onset and evolution of transonic buffet.

\subsubsection{Shock Buffeting}

In the transonic regime and in the absence of external excitation,
the flow over an airfoil governed by~\eqref{eq:URANSODE} is typically observed in either steady ($\mathbf{q}(t)=\mathbf{0}$, for all $t$) or periodic ($\mathbf{q}(t)=\mathbf{q}(t+T)$, for some oscillation period $T$), depending on the combination of Mach number and angle of attack. In the latter case, the flow exhibits self-sustained oscillations characterized by the synchronized motion of the shock wave and the separated boundary layer, commonly referred to as transonic shock buffet.

\begin{figure}[h]
        \centering
        \includegraphics[width=0.8\linewidth]{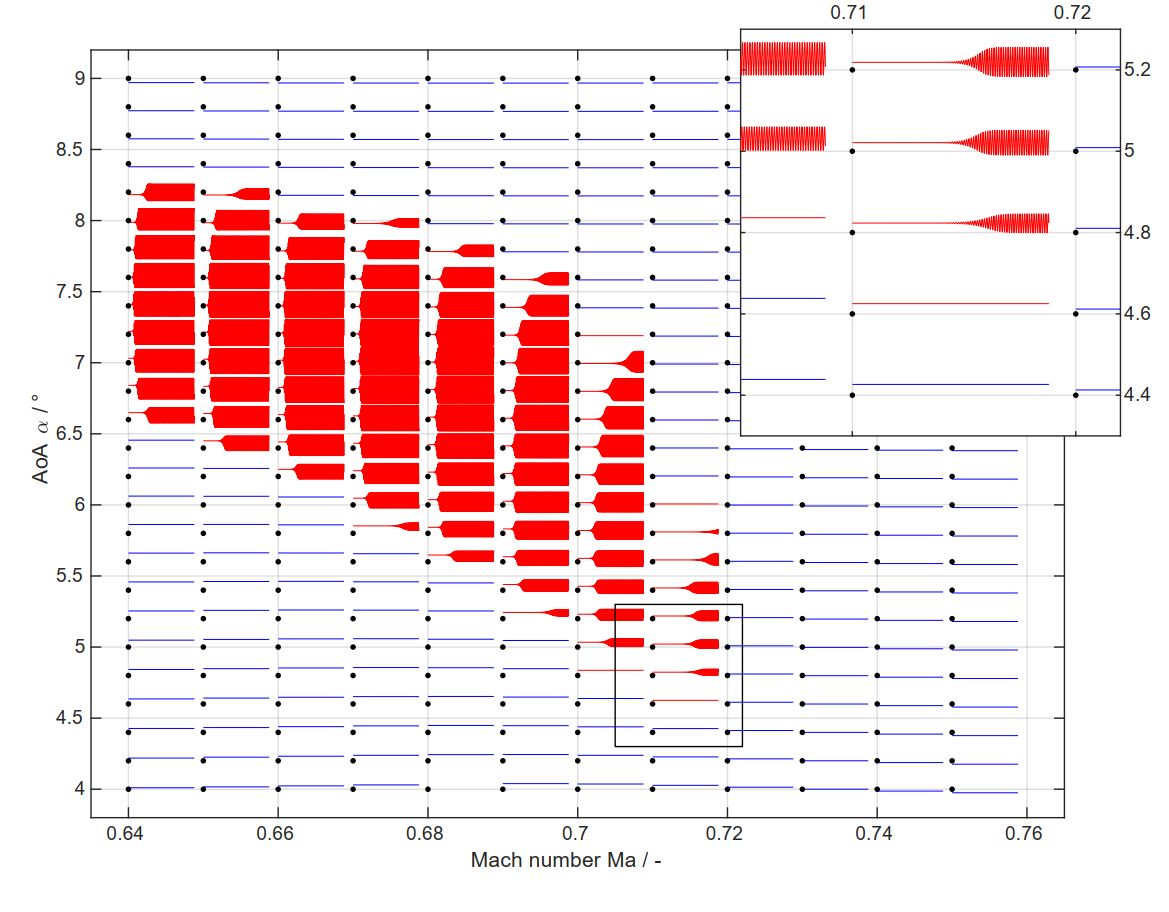}
       \caption{Unsteady lift coefficient $C_l'$ signals for different Mach numbers and angles of attack of the OAT15A airfoil, computed with the SA-neg turbulence model in~\cite{nitzsche2022effect}. Red curves correspond to cases exhibiting buffet oscillations, whereas blue curves indicate steady cases without self-sustained unsteadiness. }
       \label{fig:Nitzsche boundary} 
   \end{figure}

For the analysis that follows, we fix the Mach number $M_{\infty}$ (within the range where buffet occurs) and consider the angle of attack $\alpha$ as the governing parameter. As $\alpha$ increases, the flow transitions from a steady base flow to periodic oscillations at a critical value $\alpha_c$, marking the onset of buffeting. This is illustrated in Figure~\ref{fig:Nitzsche boundary}, showing the unsteady lift coefficient $C_l'$ for OAT15A airfoil. The buffet is observed for $M_{\infty}=0.64-0.72$ with the onset corresponding to the emergence of nonzero oscillation amplitude in $C_l'$. The critical angle of attack $\alpha_c$ decreases with increasing $M_{\infty}$.
As $\alpha$ increases further, the flow eventually restabilizes and returns to a steady state, referred to as the \textit{buffet offset}.

\subsubsection{Hopf bifurcation}

The onset of 2D transonic buffet instability, as the angle of attack $\alpha$ crosses a critical value $\alpha_c$ at fixed $M_\infty$, is associated with a codimension-one Hopf bifurcation (\cite{crouch2009origin,sartor2015stability}). In this scenario, a previously stable equilibrium $\mathbf{q}_b$ loses stability through a single complex-conjugate eigenvalue pair crossing the imaginary axis, while all remaining eigenvalues retain negative real parts (\cite{guckenheimer2013nonlinear}). The associated unstable complex-conjugate eigenvector pair 
is commonly referred to as the \textit{global mode}.
This leads to the emergence of self-sustained oscillations which settle on a limit cycle.

Depending on the nonlinear saturation mechanism, a Hopf bifurcation may be either supercritical or subcritical (\cite{kuznetsov1998elements}). In the supercritical case, a stable limit cycle of small amplitude emerges smoothly beyond the critical point, whereas a subcritical bifurcation is characterized by an abrupt loss of stability due to the presence of an unstable limit cycle. For the OAT15A airfoil, the buffet bifurcation has been shown to be supercritical in~\cite{crouch2024weakly}, consistent with prior observations in~\cite{sansica2022system} and with the smooth growth of the lift coefficient fluctuations $C_l'$ shown in Figure~\ref{fig:Nitzsche boundary}.

\subsubsection{Invariant Manifold}

For a dynamical system satisfying the conditions of a Hopf bifurcation, the Center Manifold Theorem (see \cite{kuznetsov1998elements}) guarantees the existence of a two-dimensional, parameter-dependent local invariant manifold $\mathcal{W}_{\alpha}$ for sufficiently small deviations of the bifurcation parameter $\alpha$ from its critical value. For a fixed $\alpha$, slightly above the critical point in a supercritical Hopf bifurcation, $\mathcal{W}_{\alpha}$ coincides with the unique unstable manifold of the equilibrium, hereafter denoted simply as $\mathcal{W}$, until the system reaches the stable limit cycle, beyond which uniqueness is lost (\cite{kuznetsov1998elements}). The manifold $\mathcal{W}$ is tangent at the unstable equilibrium to the plane spanned by the global mode 
referred to as the \textit{tangent plane}. Since all remaining eigenvalues have negative real parts, $\mathcal{W}$ is exponentially attracting (\textit{normally hyperbolic}), implying that trajectories in the high-dimensional state space are exponentially drawn toward it. The dynamics restricted to this manifold are topologically equivalent to the Stuart--Landau equation, which constitutes the normal-form of the Hopf bifurcation.

Thus, the long-term flow dynamics governing the 2D buffet instability evolve on the invariant manifold $\mathcal{W}$. As discussed above, close to buffet onset the limit cycle is guaranteed to lie on this manifold. We assume that this remains the case beyond onset, and verify this assumption numerically. This invariant manifold therefore captures the essential low-dimensional geometry governing the system’s asymptotic behavior.

\subsection{Scope and Objectives}

The transonic buffet dynamics governed by~\eqref{eq:URANSODE} constitute a high-dimensional nonlinear system with $\mathcal{O}(10^6)$ degrees of freedom, rendering real-time prediction and control design computationally prohibitive and motivating the need for reduced-order representations. The reduced-order modeling of this flow is challenged by strong compressibility effects and shock-induced gradients, as well as by the coexistence of multiple attractors, namely the steady base flow and the nonlinear limit cycle.

The existence of a two-dimensional invariant manifold associated with the Hopf bifurcation nevertheless provides a natural foundation for nonlinear model reduction. The objective of the present work is to leverage this invariant-manifold structure to construct a reduced-order model for transonic buffet that (i) captures the nonlinear dynamics, (ii) is physically interpretable, and (iii) predicts and reconstructs the full flow state across the phase space, including the vicinity of the unstable equilibrium, the nonlinear transient evolution, and the saturated limit cycle.

To this end, we propose an invariant-manifold-based modeling approach that is computationally efficient for extremely high-dimensional systems and does not require exhaustive sampling of the phase space. The methodology is detailed in the following section.

\section{Reduced-order modeling}\label{sec:ROM}

In this section, we present an adaptation of the data-driven invariant-manifold-based framework of~\cite{cenedese2022data} for constructing reduced-order models. The framework is formulated for fixed system parameters and therefore yields non-parametric reduced-order models that represent the dynamics associated with a given operating condition. While the methodology is general, we focus on systems undergoing a Hopf bifurcation, which is directly relevant to transonic buffet dynamics. The proposed algorithm is well suited for large-scale CFD applications and remains effective even when samples near the unstable equilibrium are limited or unavailable, without requiring prior knowledge of suitable reduced coordinates.

The remainder of this section is organized as follows. 
Section~\ref{ROMsInvariantManifold} establishes the general theoretical framework for invariant-manifold–based reduced-order modeling.
Section~\ref{Identification} discusses the data-driven formulation enabling manifold and dynamics identification from data, without requiring the knowledge of governing equations.  
Section~\ref{ProposedAlgorithm} presents the resulting algorithmic workflow, including manifold and reduced dynamics identification and extended normal-form transformation. 
Finally, Section~\ref{HopfInterpretation} provides interpretation for systems undergoing Hopf bifurcation and demonstrates how extended normal-forms produce physically meaningful modal representations.

\subsection{ROMs using Invariant Manifolds}\label{ROMsInvariantManifold}

Here we assume the existence of an attracting, normally hyperbolic $r$-dimensional \emph{invariant manifold} 
$\mathcal W \subset \mathbb R^n$ attached to the equilibrium of the system, which is globally parameterized by reduced coordinates
$\boldsymbol{\eta}\in\mathbb R^r$ with $r\ll n$.  
Following the terminology of~\cite{szalai2023data},
adapted to continuous-time dynamical systems, we refer to the parameterization
$\mathbf W:\mathbb R^r\to\mathbb R^n$, with image 
\[
\mathcal W \equiv \mathbf W(\mathbb R^r),
\]
as a \emph{decoder}.
The dynamics on $\mathcal W$ are described in the reduced coordinates by the low-dimensional system
\begin{equation}\label{eq:ROM}
\dot{\boldsymbol{\eta}} = \mathbf r(\boldsymbol{\eta}),
\end{equation}
where $\mathbf r:\mathbb R^r\to\mathbb R^r$ is the reduced vector field. Let $\Phi_t$ and $S_t$ denote the flow maps of the full-order system
\eqref{eq:URANSODE} and the reduced system~\eqref{eq:ROM}, respectively, which we
assume are well defined for all times. The requirement that
the reduced dynamics~\eqref{eq:ROM} represent the full-order dynamics on the
invariant manifold $\mathcal W$ is expressed by the following trajectory-pairing
relation: for any initial conditions $\boldsymbol{\eta}_0 \in \mathbb R^r$ and
$\mathbf q_0 := \mathbf W(\boldsymbol{\eta}_0) \in \mathcal W$,
\begin{equation}\label{eq:trajectory}
\Phi_t(\mathbf q_0)
=
\mathbf W(S_t(\boldsymbol{\eta}_0)),
\qquad \text{for all } t .
\end{equation}
According to the left-hand side of~\eqref{eq:trajectory}, any trajectory
$\boldsymbol{\eta}(t)=S_t(\boldsymbol{\eta}_0)$ of the reduced-order system is
mapped to a trajectory
$\mathbf q(t)=\Phi_t(\mathbf q_0)$ of the full-order system, which
remains on the invariant manifold $\mathcal W$ (the image of $\mathbf W$) for all
times. We call a pair $(\mathbf r,\mathbf W)$ satisfying this condition a reduced-order
model (ROM) of~\eqref{eq:URANSODE}. Taking the time-derivative of~\eqref{eq:trajectory} yields the \emph{invariance equation}
\begin{equation}\label{eq:invariance}
\mathrm D\mathbf W(\boldsymbol{\eta})\,\mathbf r(\boldsymbol{\eta})
=
\mathbf f\!\left(\mathbf W(\boldsymbol{\eta})\right),
\qquad \text{for all } \boldsymbol{\eta}\in\mathbb R^r .
\end{equation}

We consider parameterizations satisfying $\mathbf W(\mathbf 0)=\mathbf 0$, so that the equilibrium
$\boldsymbol{\eta}=\mathbf 0$ of~\eqref{eq:ROM} corresponds to the equilibrium $\mathbf q=\mathbf 0$
of~\eqref{eq:URANSODE}. The tangent space to $\mathcal W$ at $\mathbf q=\mathbf 0$, denoted
$T_{\mathbf 0}\mathcal W$ and spanned by the columns of $\mathrm D\mathbf W(\mathbf 0)$, defines an
$r$-dimensional invariant subspace of the linearized full-order dynamics $\mathrm D\mathbf f(\mathbf 0)$.
For systems that are stable at the origin and undergo a supercritical Hopf bifurcation, this subspace
coincides with the two-dimensional unstable eigenspace of the linearization.

The main idea behind the parameterization method to determine invariant manifolds and their reduced dynamics is to consider Taylor expansions of both $\mathbf W(\boldsymbol{\eta})$ and $\mathbf r(\boldsymbol{\eta})$ in the parameters, and to solve the invariance equation at successive orders by collecting terms of the same order in~\eqref{eq:invariance}. This yields a sequence of cohomological equations that \emph{simultaneously} determine the coefficients of $\mathbf W(\boldsymbol{\eta})$ and $\mathbf r(\boldsymbol{\eta})$ at each order (see details in~\cite{haro2016parameterization}). The manifold parameterization is an underdetermined problem, which is further reflected in the fact that the cohomological equations admit infinitely many solutions. This flexibility allows for the choice of different \emph{parameterization styles}, among which the \emph{graph style} and the \emph{normal-form style} are the most common.

In the graph-style parameterization, the invariant manifold is represented as a
graph over the $r$-dimensional tangent space $T_{\mathbf 0}\mathcal W$, with a
complementary subspace $V$, i.e., such that $\mathbb R^n = T_{\mathbf 0}\mathcal W \oplus V$. In this
case, the manifold can be written as
\[
\mathcal W
=
\{(\boldsymbol q_1,\boldsymbol q_2)\in T_{\mathbf 0}\mathcal W\times V
\;|\;
\boldsymbol q_2=\widetilde{\boldsymbol g}(\boldsymbol q_1)\},
\]
for some nonlinear function $\widetilde{\boldsymbol g}:T_{\mathbf 0}\mathcal W\to V$. After
choosing bases of $T_{\mathbf 0}\mathcal W$ and $V$, this representation induces a parameterization
of the form
\begin{align} \label{eq:graphstyle}
\mathbf W(\boldsymbol{\eta})
=\mathbf{B}[\boldsymbol{\eta}\;\;\boldsymbol{\widehat g}(\boldsymbol{\eta})]^\top
\equiv \mathbf{W}_1\boldsymbol{\eta}+\boldsymbol g(\boldsymbol{\eta}).
\end{align}
Here $\mathbf{B}$ is the coordinate transformation from the chosen basis of $T_{\mathbf 0}\mathcal W\oplus V$ to the canonical basis of $\mathbb R^n$, $\mathbf{W}_1$ comprises the first $r$ columns of $\mathbf{B}$, which span the tangent space $T_{\mathbf 0}\mathcal W$, and $\boldsymbol g:\mathbb R^r\to\mathbb R^n$ is the nonlinear part of the parameterization.

In the normal-form style, the invariance equation is solved in coordinates that transform the reduced dynamics $\mathbf r$ into a classical normal-form,
systematically eliminating non-essential nonlinear terms.

Solving the invariance equation directly is computationally challenging for systems of high dimension. While approaches exist to reduce this effort in~\cite{jain2022compute}, in the present work we adopt a data-driven approach that avoids explicit solution of the invariance equation and relies solely on discrete-time measurements of the full state $\mathbf{q}$. This approach has the added advantage of being applicable to data from diverse sources, including experiments, without requiring direct access to the governing equations.

  \subsection{Data-Driven Manifold and Reduced Dynamics} 
\label{Identification}

We now consider identification of a decoder $\mathbf{W}(\boldsymbol{\eta})$ and
reduced dynamics $\mathbf{r}(\boldsymbol{\eta})$ from sampled data $\mathbf{q}_1, \hdots , \mathbf{q}_N$ of the full-order system~\eqref{eq:URANSODE} at successive time instances
$t_1,\ldots,t_N$\footnote{The consideration of a single trajectory is only used for notational convenience. The presented approach directly generalizes to the case in which data are collected from multiple trajectories.}. Identifying these maps amounts to selecting finite-dimensional
(typically polynomial) representations for $\mathbf W$ and $\mathbf r$ such that
the sampled data are consistent with the invariance equation~\eqref{eq:invariance}
when expressed in the reduced coordinates
$\boldsymbol{\eta}_k = \mathbf W^{-1}(\mathbf q_k)$.
Here, $\mathbf W^{-1}$ denotes the inverse of the parameterization map $\mathbf W$, which is 
defined on the invariant manifold $\mathcal W$.

It is convenient to decompose the
identification problem into two stages: first, identifying a suitable
parameterization $\mathbf W$ of a candidate manifold in the state space, and
subsequently determining the reduced dynamics $\mathbf r$ on that manifold.

To formalize this idea, we denote by $\mu \mapsto \mathbf W_\mu$ a family of
candidate parameterizations 
and assume that for some parameter value $\mu_\star$,
the map $\mathbf W_{\mu_\star}$ yields an exact parameterization of the true
invariant manifold $\mathcal W$, that is,
\begin{align}
\mathcal W = \mathcal W_{\mu_\star},
\qquad
\mathcal W_\mu := \mathbf W_\mu(\mathbb R^r).
\end{align}

\begin{figure}[htbp]
  \centering
  \includegraphics[
    width=0.9\linewidth,
    trim=20pt 550pt 20pt 50pt, 
    clip
  ]{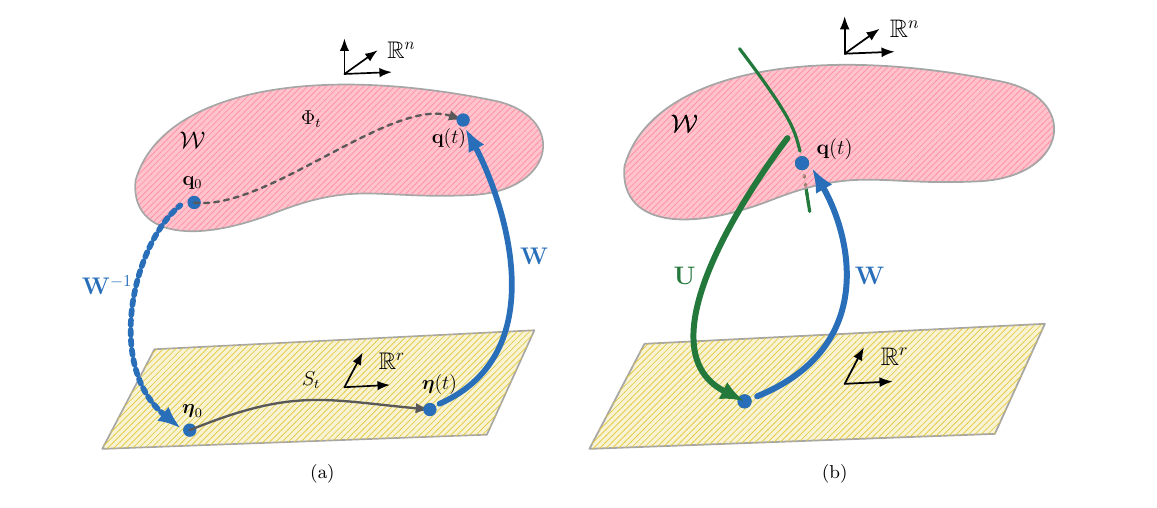}
\caption{Invariant-manifold formulation versus autoencoder formulation.
(a) \emph{Invariant-manifold setting.} An attracting, normally hyperbolic invariant
manifold $\mathcal W \subset \mathbb R^n$ is globally parameterized by reduced
coordinates $\boldsymbol{\eta}\in\mathbb R^r$ through a decoder (embedding)
$\mathbf W:\mathbb R^r\to\mathbb R^n$, with $\mathcal W=\mathbf W(\mathbb R^r)$.
The reduced dynamics $\dot{\boldsymbol{\eta}}=\mathbf r(\boldsymbol{\eta})$ and
the embedding $\mathbf W$ satisfy the invariance equation
which ensures trajectory pairing
$\Phi_t(\mathbf q_0)=\mathbf W(S_t(\boldsymbol{\eta}_0))$.
The inverse map $\mathbf W^{-1}$ is defined only on the manifold $\mathcal W$.
(b) \emph{Autoencoder (data-driven) setting.} An encoder
$\mathbf U:\mathbb R^n\to\mathbb R^r$ is introduced, defined on the entire ambient
space $\mathbb R^n$, extending $\mathbf W^{-1}$ away from $\mathcal W$.
Together with the decoder $\mathbf W$, the pair $(\mathbf U,\mathbf W)$ is
characterized by the consistency relations
$\mathbf U\circ\mathbf W=\mathrm{Id}$ and, approximately on data,
$\mathbf W\circ\mathbf U\approx \mathrm{Id}$, enabling identification of a
candidate manifold from sampled trajectories.}
\label{fig:invariant_vs_autoencoder}
\end{figure}

With this notation, the data-driven identification problem reduces to finding
parameter values $\mu$ for which the sampled data lie on the corresponding
candidate manifold, namely,
\begin{align}
\mathbf q_k \in \mathcal W_\mu
\quad \text{for all} \quad k=1,\ldots,N.
\end{align}
To express this inclusion in terms of algebraic conditions, we introduce for
each $\mu$ a map $\mathbf U_\mu:\mathbb R^n \to \mathbb R^r$ that assigns reduced
coordinates to full-state data points. We assume that $\mathbf U_\mu$ extends the
inverse $\mathbf W_\mu^{-1}$ from the manifold $\mathcal W_\mu$ to the ambient
state space $\mathbb R^n$, and we refer to $\mathbf U_\mu$ as an \emph{encoder}.
This extension is necessary because the inverse map $\mathbf W_\mu^{-1}$ is only
defined on $\mathcal W_\mu$, and for generic parameter values $\mu$ the data
$\mathbf q_k$ will not lie exactly on this set. Under this construction, the requirement $\mathbf q_k \in \mathcal W_\mu$ is
equivalent to the reconstruction condition
\begin{align}\label{autoencoder:data:driven}
\mathbf W_\mu (\mathbf U_\mu(\mathbf q_k))
=
\mathbf q_k,
\qquad \text{for all} \quad k=1,\ldots,N,
\end{align}
which enforces that encoding followed by decoding reproduces the observed data.
This relation has the form of an autoencoder consistency condition. Specifically, we refer to $(\mathbf{U}_\mu,\mathbf W_\mu)$ as an \textit{autoencoder} pair, which is characterized by the condition $\mathbf U_\mu\circ~\mathbf W_\mu=~{\rm Id}$. The distinction between the classical invariant-manifold formulation and the
data-driven autoencoder viewpoint introduced below is illustrated schematically
in Figure~\ref{fig:invariant_vs_autoencoder}.

The autoencoder condition is rarely satisfied in practice because (i) the finite-dimensional class of candidate manifold parameterizations (e.g., fixed-order polynomial maps) cannot exactly model the invariant manifold, and, (ii)  the data are collected by simulated or experimental trajectories, which although attracted by the manifold are not exactly initiated from it. A standard approach to resolve this issue is to  impose~\eqref{autoencoder:data:driven} approximately (see \cite{cenedese2022data}), 
by solving the optimization
problem\footnote{To simplify notation, we henceforth omit the dependence of the mappings $\mathbf U$, $\mathbf W$, and later $\mathbf r$ on their finite-dimensional parameterization denoted by $\mu$.}
\begin{equation}\label{eq:NLPCA}
\begin{aligned}
\underset{\mathbf{U}, \mathbf{W}}{\min}
&\sum_{k=1}^{N}
\big\|
\mathbf{q}_k - \mathbf{W}\big(\mathbf{U}(\mathbf{q}_k)\big)
\big\|^2
\\
\text{s.t.}\;&
\mathbf{U}\circ \mathbf{W} = {\rm Id},
\end{aligned}
\end{equation}
which identifies the manifold that best reconstructs the data. The
minimum of~\eqref{eq:NLPCA} will be far from zero if the data do not lie on an
$r$-dimensional invariant submanifold of the state space.

 To identify the reduced dynamics, we rewrite the trajectory pairing relation \eqref{eq:trajectory} as
\begin{equation}
\mathbf{W}^{-1}(\boldsymbol{\Phi}_t(\mathbf q_0))
=
S_t(\mathbf{W}^{-1}(\mathbf q_0)),
\qquad \text{for all } t \text{ and } \mathbf q_0\in\mathcal W .
\end{equation}
By extending $\mathbf W^{-1}$ to an encoder $\mathbf U:\mathbb R^n\to\mathbb R^r$, 
we get that
\[
\frac{d}{dt}\mathbf U(\mathbf q(t)) = \mathbf r(\mathbf U(\mathbf q(t))) 
\]
along the trajectory
$\mathbf q(t)=\Phi_t(\mathbf q_0)$.
In the data-driven setting, the time derivative along each trajectory is
approximated from sampled data. Specifically, for each sample
$\mathbf q_k = \mathbf q(t_k)$, we consider the approximation
\begin{align} \label{finite:differences}
\frac{d}{dt}\mathbf U(\mathbf q(t))\big|_{t=t_k}
\;\approx\;
\frac{\mathbf U(\mathbf q_{k+1}) - \mathbf U(\mathbf q_k)}{t_{k+1}-t_k}.
\end{align}

The reduced dynamics are then identified by solving the optimization problem 
\begin{equation}\label{eq:autodyn}
\underset{\mathbf{r}}{\min}
\sum_{k=1}^{N}
\Big\|
\frac{d}{dt}\mathbf U(\mathbf q(t))\big|_{t=t_k}
-
\mathbf r(\mathbf U(\mathbf q_k))
\Big\|^2,
\end{equation}
which yields a best-fit evolution law for the reduced coordinates along the
sampled trajectory (as already pointed out, the exact same formulation can be used to identify the reduced dynamics from multiple trajectories).

For high-dimensional systems, finding a nonlinear polynomial encoder
$\mathbf U:\mathbb R^n\to\mathbb R^r$ is often computationally intractable, as even for relatively low orders, the  number of monomials that are used to describe it blows up with the dimension $n\gg 1$ of the full state.  
This motivates seeking reduced-order
representations in which the encoder is linear, so that reduced coordinates can
be extracted efficiently from data. Such a situation arises when the invariant manifold admits a \emph{graph-style}
representation~\eqref{eq:graphstyle}. 
In this setting, a natural choice of encoder is the linear map
\[
\mathbf U(\mathbf{q})
:=
\bigl[\,\mathbf I_{r\times r}\;\;\mathbf 0\,\bigr]\,\mathbf{B}^{-1}\mathbf{q}\equiv \mathbf U_1^\top\mathbf{q},
\]
which simply extracts the coordinates of a vector in the decomposition $\mathbb R^n = T_{\mathbf 0}\mathcal W \oplus V$ along the tangent space.

\subsection{Proposed Reduced-Order modeling Approach}\label{ProposedAlgorithm}

We now present the proposed reduced-order modeling algorithm, which consists of two stages: (i) data-driven identification of the invariant manifold and its internal dynamics, and (ii) transformation of the reduced dynamics and the associated manifold parameterization into extended normal-form coordinates.

The first stage of the algorithm is sufficient to construct a reduced-order model, whereas the second stage is introduced to obtain a simpler and more physically interpretable representation.

\subsubsection{Stage I: Data-driven Manifold and Dynamics Identification}
The identification of the invariant manifold and the reduced dynamics is performed in two separate steps.

\paragraph{Manifold Identification}

To identify the invariant manifold, we seek encoder and decoder maps
$\mathbf{U}(\mathbf{q})$ and $\mathbf{W}(\boldsymbol{\eta})$ that satisfy the
autoencoder condition~\eqref{eq:NLPCA}. To make the approach tractable for
high-dimensional systems, we restrict the encoder to be linear,
\begin{equation}\label{eq:linearencoder}
    \boldsymbol{\eta} = \mathbf{U}(\mathbf{q}) := \mathbf{U}_1^\top \mathbf{q},
\end{equation}
where $\boldsymbol{\eta} \in \mathbb{R}^r$ are the reduced coordinates and
$\boldsymbol{\eta}=\mathbf{0}$ corresponds to the equilibrium. 
The decoder $\mathbf{W}(\boldsymbol{\eta})$ is expressed as an $M$th-degree
polynomial in the reduced coordinates, i.e., 
\begin{equation}\label{eq:manifoldexpansion}
    \mathbf{W}(\boldsymbol{\eta})
    =
    \sum_{i=1}^{M} \mathbf{W}_i \boldsymbol{\eta}^{\,i}
    =
    \mathbf{W}_1 \boldsymbol{\eta}
    +
    \mathbf{W}_{2:M} \boldsymbol{\eta}^{2:M},
\end{equation}
where $\boldsymbol{\eta}^{\,i}$ collects all monomials of total degree $i$ and
$\mathbf{W}_i \in \mathbb{R}^{n\times K_i}$ are the corresponding coefficient
blocks. The stacked notation
\[
    \mathbf{W}_{2:M}
    =
    [\mathbf{W}_2 \;\mathbf{W}_3 \;\cdots\; \mathbf{W}_M], \qquad
    \boldsymbol{\eta}^{2:M}
    =
    [\boldsymbol{\eta}^2\;\;\boldsymbol{\eta}^3\;\;\cdots\;\;\boldsymbol{\eta}^M]^\top
\]
compactly groups all nonlinear contributions. For instance, if
$\boldsymbol{\eta}=[\eta_1\;\eta_2]^\top$, then
$\boldsymbol{\eta}^{\,i}=[\eta_1^i\;\;\eta_1^{i-1}\eta_2\;\;\cdots\;\;\eta_2^i]$.

Substituting~\eqref{eq:linearencoder} and~\eqref{eq:manifoldexpansion} into the
autoencoder identification cost~\eqref{eq:NLPCA} yields the manifold-identification problem 
\begin{subequations} \label{data:driven:linear:autoencoder}
\begin{equation}\label{eq:nonconvex}
\begin{aligned}
\min_{\mathbf{U}_1,\mathbf{W}_1,\mathbf{W}_{2:M}}
\;&
\sum_{k=1}^{N}
\big\|
\mathbf{q}_k
-
\mathbf{W}_1 \mathbf{U}_1^\top \mathbf{q}_k
-
\mathbf{W}_{2:M}(\mathbf{U}_1^\top \mathbf{q}_k)^{2:M}
\big\|^2
\end{aligned}
\end{equation}
subject to the constraints 
\begin{align} \label{eq:nonconvex_constraints}
\mathbf{U}_1^\top \mathbf{W}_1 = \mathbf{I}\quad {\rm and}\quad \mathbf{U}_1^\top \mathbf{W}_{2:M} = \mathbf{0}. 
\end{align}
\end{subequations}
This corresponds to a data-driven identification of the graph-style parameterization~\eqref{eq:graphstyle}, in which the tangent space $T_{\mathbf 0}\mathcal W$  is spanned by the columns of $\mathbf{W}_1$, and the complementary subspace $V$ is the orthogonal complement of the space spanned by the columns of $\mathbf{U}_1$.

The optimization problem~\eqref{data:driven:linear:autoencoder} is generally nonconvex due to the bilinear coupling between $\mathbf{U}_1$ and $\mathbf{W}_1$, as well as the dependence of the higher-order terms on $\mathbf{U}_1$. Thus, directly solving~\eqref{data:driven:linear:autoencoder}, as done in~\cite{cenedese2022data}, becomes computationally prohibitive for very high-dimensional systems.

A substantial simplification occurs when the matrix $\mathbf{U}_1$ of a suitable reduced-coordinate map is known \emph{a priori}. This situation can arise in two common ways: (i) the tangent space at the equilibrium is known, in which case one fixes the columns of $\mathbf{W}_1$ to the orthonormal basis of it and chooses the encoder $\mathbf{U}_1^\top \mathbf{q} := \mathbf{W}_1^\top \mathbf{q}$; or, more generally, (ii) another suitable $r$-dimensional basis is chosen to determine the range of the encoder, comprising the columns of $\mathbf{U}_1$. In both cases, once $\mathbf{U}_1$ is fixed, the remaining coefficients are identified as the unique solution of the constrained least-squares problem. According to the constraints of this problem, the columns of $\mathbf{U}_1$ are biorthogonal to those of $\mathbf{W}_1$ and annihilate the columns of $\mathbf{W}_{2:M}$.

Such prior knowledge of the encoder typically relies on the data being close to equilibrium (\cite{bettini2025data,buurmeijer2025taming}) or on having suitable reduced coordinates. The latter was exploited for the cylinder wake in~\cite{cenedese2022data}, where the most energetic POD modes on the limit cycle are known to provide a suitable reduced basis (\cite{noack2003hierarchy}). However, this information is generally unavailable for high-dimensional flow problems such as shock buffeting. 

Here, we focus on problems where the manifold does not fold over its tangent space when viewed as a graph over it with values in its orthogonal complement. In these cases, an encoder as in case (i) above is well-suited. By contrast, an arbitrary $r$-dimensional subspace for constructing an encoder can be unsuitable. Specifically, projecting along the orthogonal complement of such a subspace can lead to two issues: (i) the distance between the data and their projections onto the manifold (i.e., points on the manifold with the same projection onto the subspace) may be much larger than the true distance to the manifold, or (ii) the projection may fail to be one-to-one due to folds in the manifold when viewed as a graph over the chosen subspace.    

To overcome these limitations, we introduce an iterative update of the encoder, summarized in Algorithm~\ref{alg:tangent_refinement}. The first step of the algorithm identifies the manifold using a fixed linear encoder $\mathbf{U}_1$, producing a data-driven parameterization based on prescribed reduced coordinates. Subsequent iterations update the encoder using the linear part of the identified parameterization, until the process converges to an orthogonal graph-style parameterization.

\vspace{1em}

\begin{algorithm}[H]
\fbox{
\parbox{0.9\linewidth}{
\caption{Iterative update of linear encoder and manifold coefficients}
\label{alg:tangent_refinement}
\begin{enumerate}
  \item \textbf{Initialization.}
  Choose an initial encoder matrix $\mathbf{U}_1^{(0)}\in\mathbb{R}^{n\times r}$
  with orthonormal columns. When data near the equilibrium are unavailable, a
  suitable initial choice is given by leading POD or DMD modes computed from data
  on an attracting solution.

  \item \textbf{Iterative update.}
  For $i = 0,1,2,\dots$, compute the reduced coordinates
$\boldsymbol{\eta}_k^{(i)}=\mathbf{U}_1^{(i)\top}\mathbf{q}_k$ and solve the \textit{convex} optimization problem
  \begin{equation}
    (\mathbf{W}_1^{(i+1)}, \mathbf{W}_{2:M}^{(i+1)})
    =
    \underset{\mathbf{W}_1,\,\mathbf{W}_{2:M}}{\argmin}   
    \sum_{k=1}^{N}
    \big\|
      \mathbf{q}_k
      -
      \mathbf{W}_1\,\boldsymbol{\eta}_k^{(i)}
      -
      \mathbf{W}_{2:M}(\boldsymbol{\eta}_k^{(i)})^{2:M}
    \big\|^2 .
  \end{equation}

  \item \textbf{Orthonormalization.}
  Update the encoder matrix via
  \[
  \mathbf{U}_1^{(i+1)} \leftarrow
  \operatorname{orth}(\mathbf{W}_1^{(i+1)}),
  \]
 where $\operatorname{orth}(\mathbf{W}_1^{(i+1)})$ denotes an orthonormalization of the matrix $\mathbf{W}_1^{(i+1)}$.  
  \item \textbf{Convergence check.}
  Stop when
  \[
  \big\|\mathbf{W}_1^{(i+1)} - \mathbf{W}_1^{(i)}\big\| < \varepsilon,
  \]
  where $\varepsilon$ is a prescribed tolerance.
\end{enumerate}
}
}
\end{algorithm}

\vspace{1em}

By exploiting the Kronecker structure of the monomial expansion, the minimization problem at each
iteration reduces to a linear least-squares system whose unknowns correspond only to the polynomial
coefficients. Each update can therefore be computed efficiently using the pseudoinverse of an
$N\times K$ matrix (see Appendix~\ref{appA}), where $K$ is the number of distinct monomials of total degree up to $M$ in $r$ variables,  rather than operating in the full $n$-dimensional state space. 
Consequently, each iteration remains computationally tractable even for very high-dimensional
datasets, such as URANS state vectors with $n=\mathcal{O}(10^6)$. A small number of iterations is
typically sufficient to reduce the dependence of the tangent-plane approximation on the initial
choice of coordinates. The converged representation satisfies the corresponding graph-style constraints~\eqref{eq:nonconvex_constraints}.

\paragraph{Dynamics Identification}
The internal dynamics on the manifold are modeled by a polynomial vector field of order~$P$:
\begin{equation}\label{eq:S}
    \mathbf{r}(\boldsymbol{\eta}) = \mathbf{L} \boldsymbol{\eta} + \mathbf{R}_{2:P} \boldsymbol{\eta}^{2:P},
\end{equation}
where \( \mathbf{L} \) denotes the linear dynamics near the equilibrium and \( \mathbf{R}_{2:P} \) collects the higher-order coefficients
associated with all monomials of the reduced variables up to degree~\( P \). 
The coefficients of the reduced vector field \( \mathbf{r}(\boldsymbol{\eta}) \) are identified by solving~\eqref{eq:autodyn} with $\mathbf{U}=\mathbf{U}_1$ and $\mathbf{r}$ given, which reduces the identification to a linear least-squares problem in the coefficients of~\( \mathbf{r} \).
For sufficiently small time steps, the time derivative
\( d(\mathbf{U}_1^\top \boldsymbol q(t))/dt|_{t=t_k} \)
in~\eqref{eq:autodyn} can be approximated using finite differences as in~\eqref{finite:differences}.

\subsubsection{Stage II: Extended Normal-Form Transform}
\label{sec:extendednormalform}

The second stage transforms the reduced dynamics into the so-called \textit{extended normal-form} introduced for data-driven invariant manifold identification in~\cite{cenedese2022data}. 
The corresponding coordinate change, in turn, yields an updated manifold expansion. Together, the two stages constitute a data-driven counterpart of the classical normal-form manifold parameterization.
 This stage is optional and can be applied when the user seeks a simpler, more interpretable representation of the ROM. 

Starting from the identified reduced dynamics~\eqref{eq:S}, we assume that their Jacobian $\mathbf{L}$ is semisimple and thus diagonalizable. Therefore, there exists a transformation matrix $\mathbf{T} \in \mathbb{C}^{r \times r}$ such that $\mathbf{L}\mathbf{T} = \mathbf{T}\boldsymbol{\Lambda}$, where $\boldsymbol{\Lambda} = \mathrm{diag}(\lambda_1, \ldots, \lambda_r)$ is the diagonal matrix with corresponding  entries the eigenvalues of $\mathbf{L}$. 

In classical normal-form theory, the dynamics are transformed into a simpler form by a near-identity transformation that is determined through a sequence of cohomological equations. 
In particular, we consider the nonlinear change of variables
\begin{equation}\label{eq:h_transform}
    \boldsymbol{\eta} = \mathbf{T}\ \circ \mathbf{h}(\mathbf{z}) 
    = \mathbf{T}\mathbf{z} + \mathbf{T}\,\mathbf{H}_{2:Q}\,\mathbf{z}^{2:Q},
    \qquad  
    \mathbf{z} = \mathbf{h}^{-1} \circ (\mathbf{T}^{-1} \boldsymbol{\eta})
    = \mathbf{T}^{-1} \boldsymbol{\eta} + \mathbf{H}^{-1}_{2:Q}(\mathbf{T}^{-1}\boldsymbol{\eta})^{2:Q},
\end{equation}
where $\mathbf{T}\in\mathbb{C}^{r\times r}$ diagonalizes the linear part of the dynamics and
$\mathbf{h}$ is a near-identity polynomial mapping of order $Q$,
\[
\mathbf{h}(\mathbf{z}) = \mathbf{z} + \mathbf{H}_{2:Q}\,\mathbf{z}^{2:Q}.
\]
Here, $\mathbf{H}_{2:Q}$ collects the monomial coefficients of degrees
$2$ through $Q$ in $\mathbf{z}$, and $\mathbf{h}^{-1}$ is the inverse of
$\mathbf{h}$, whose expansion to each order is obtained by the requirement
\[
\mathbf{h}^{-1}\circ \mathbf{h}(\mathbf{z}) = \mathbf{z} + \mathcal{O}(\|\mathbf{z}\|^{Q+1}).
\]

Taking the time derivative of the reduced trajectory in the transformed coordinates, we obtain from~\eqref{eq:h_transform} and~\eqref{eq:S} a homological
equation at each order that determines the corresponding coefficients of $\mathbf{H}_{2:Q}$ and the remaining nonlinear terms in the transformed vector field (see, e.g.,~\cite{murdock2003normal}). 
The resulting transformed dynamics take the form
\begin{equation}\label{eq:normalform}
   \frac{d}{dt} \mathbf{z}= \boldsymbol{\Lambda}\mathbf{z} 
   + \mathbf{N}_{2:Q}\,\mathbf{z}^{2:Q},
\end{equation}
truncated at order $Q$, where the linear part is diagonal. 

The cohomological equations successively annihilate higher-order monomial terms in the dynamics under some non-resonance conditions that are exclusively related to the linear part $\mathbf{L}$.
Since generic hyperbolic systems of the same stability type meet those non-resonance requirements, their dynamics can be fully linearized by Poincare's theorem. However, the region of validity of the respective transformations may be severely restricted, as an exact linearization excludes the possibility of e.g., capturing limit cycles, which appear at the vicinity of the origin near a supercritical Hopf bifurcation.    

To resolve such issues, the extended normal-form transformations of~\cite{cenedese2022data} deliberately retain  \emph{near-resonant} terms. This excludes small denominators from the cohomological equation, which would otherwise yield excessively large coefficients in the nonlinear part of the transformation, thereby constraining its effective domain to a narrow neighborhood of the fixed point. 
Once the reduced dynamics are transformed into their (extended) normal-form, the invariant
manifold transforms accordingly via the composition 
\begin{equation}\label{eq:manifoldnormal}
    \tilde{\mathbf{W}}(\mathbf{z}) 
    :=\mathbf{W}(\mathbf{T}\ \circ \mathbf{h}(\mathbf{z})),
\end{equation}
which yields the expansion
\begin{equation}
    \tilde{\mathbf{W}}(\mathbf{z}) 
    = \tilde{\mathbf{W}}_1 \mathbf{z} + \tilde{\mathbf{W}}_{2:(M\cdot Q)} \mathbf{z}^{2:(M\cdot Q)}.
\end{equation}

Since the transformed reduced dynamics have a diagonal
linear part, the columns of $\tilde{\mathbf{W}}_1$ represent the eigenvectors of the
Jacobian of~\eqref{eq:URANSODE}, which span the tangent plane to the manifold at the equilibrium.

\subsection{Normal-form structure and physical interpretation for Hopf bifurcations} \label{HopfInterpretation}

For systems undergoing a Hopf bifurcation, the extended normal-form~\eqref{eq:normalform}
reduces to a single complex equation.
Introducing polar coordinates $z = r e^{\imath\phi}$ yields the decoupled amplitude–phase
representation
\begin{equation}\label{eq:hopf_polar_general}
\dot{r} = \sigma r + \sum_{i \in \mathcal{R}_r} \beta_i r^i,
\qquad
\dot{\phi} = \omega + \sum_{i \in \mathcal{R}_\phi} \delta_i r^i,
\end{equation}
where $\sigma$ and $\omega$ are the real and imaginary parts of the eigenvalue
$\lambda = \sigma + \imath\omega$, and the index sets of retained resonant orders are
\[
\mathcal{R}_r = \{3, 5, 7,\dots\}, \qquad
\mathcal{R}_\phi = \{2, 4, 6,\dots\}.
\]
The coefficients $\beta_i$ govern the nonlinear amplitude dynamics, while the coefficients $\delta_i$ are nonlinear corrections of the oscillation frequency.

The cubic truncation of~\eqref{eq:hopf_polar_general} yields the classical
Stuart--Landau equation
\begin{equation}\label{eq:SL_cubic}
\dot{r} = \sigma r + \beta_3 r^3,
\qquad
\dot{\phi} = \omega + \delta_2 r^2,
\end{equation}
which already captures the essential nonlinear saturation mechanism.
With $\sigma/\beta_3<0$, the limit-cycle amplitude $r_\star$ is 
the non-trivial steady amplitude of~\eqref{eq:SL_cubic}, namely, 
\begin{equation}
0 = \sigma r_\star + \beta_3 r_\star^3
\quad\Rightarrow\quad
r_\star^2 = -\frac{\sigma}{\beta_3}.
\end{equation}
Its corresponding oscillation frequency is
\[
\Omega_\star = \omega + \delta_2 r_\star^2.
\]
The transient amplitude evolution admits the closed-form solution  
\begin{equation}\label{eq:r_solution}
r(t)
= \sqrt{\frac{r_\star^2}{1 + \Bigl(\dfrac{r_\star^2}{r_0^2} - 1\Bigr)\mathrm{e}^{-2\sigma t}}},
\end{equation}
where $r_0 = r(0)$ is the initial amplitude. 
By integrating the frequency equation in~\eqref{eq:SL_cubic} we also obtain the closed-form expression \begin{equation}\label{eq:phi_solution_closed}
\phi(t)
= \phi_0
+ \bigl(\omega + \delta_2 r_\star^2\bigr)t
+ \frac{\delta_2 r_\star^2}{2\sigma}
\ln\!\left(
\frac{1 + \bigl(\frac{r_\star^2}{r_0^2} - 1\bigr)\mathrm{e}^{-2\sigma t}}
     {\frac{r_\star^2}{r_0^2}}
\right)
\end{equation}
for the phase. Hence, the solution $z(t) = r(t)\,\mathrm{e}^{\imath\phi(t)}$ of the Stuart--Landau equation is available in a closed form that captures the entire transient behavior from the unstable equilibrium to the limit cycle.

For higher-order models (with additional powers from 
$\mathcal{R}_r$ and~$\mathcal{R}_\phi$), the amplitude $r_\star$ of the stable limit-cycle is the smallest positive root of the polynomial equation
\[
0 = \sigma r_\star + \sum_{i\in\mathcal{R}_r} \beta_i r_\star^i,
\]
with corresponding oscillation frequency 
\[
\Omega_\star = \omega + \sum_{i\in\mathcal{R}_\phi} \delta_i r_\star^i.
\]

In a classical normal-form of the reduced dynamics away from the bifurcation point, nonlinear terms would have been eliminated, effectively linearizing the dynamics
around the hyperbolic equilibrium. By contrast, the extended normal-form, which 
deliberately retains near-resonant nonlinearities, preserves the nonlinear saturation mechanism responsible for the limit cycle. This allows the reduced-order model to capture periodic behavior even for parameter values farther from the Hopf onset.

Throughout the transient evolution to the limit cycle, both the amplitude $r(t)$ and the instantaneous frequency $\Omega(t)=\dot{\phi}(t)$ vary slowly, on a time scale much larger than the instantaneous oscillation period $2\pi/\Omega(t)$. As a result, on each short time window $[t_0-\delta t/2,\,t_0+\delta t/2]$ comprising a few periods, we may consider the approximations
\[
r(s) \approx r(t_0),
\qquad
\phi(s) \approx \phi(t_0) + \Omega(t_0)\,(s-t_0).
\]
These are justified by the frequency equation in \eqref{eq:SL_cubic}, where a zeroth order approximation of the amplitude yields a first order approximation of the phase. 
Thus, we obtain the approximation   
\[
z(s) \approx \bar{r}(t_0)\,\mathrm{e}^{\imath\phi(t_0)}\mathrm{e}^{\imath \Omega(t_0)(s-t_0)}
\]
of the trajectory during this time interval. On the limit cycle, this representation is exact and we have
\[
    z(s) = r_{\star}e^{\imath \phi (t_0)}e^{\imath \Omega_{\star}(s-t_0)},
\]
for any choice of the initial time $t_0$.

Substituting this structure into the manifold
expansion~\eqref{eq:manifoldnormal} gives
\[
    \tilde{\boldsymbol W}(z(s)) 
    \;\approx\; 
    \boldsymbol{\Phi}_0+\sum_{k=1}^{M Q} 
        \boldsymbol{\Phi}_ke^{\imath k \Omega(t_0) (s-t_0)},
\]
with
\[
   \boldsymbol{\Phi}_k 
    := \tilde{\mathbf{W}}_k \bar r(t_0)^{k} 
       e^{\imath k \phi(t_0)},
    \qquad k = 1,\ldots,M Q.
\]

 This reveals a hierarchy of \emph{harmonic modes}: a fundamental
mode $\boldsymbol{\Phi}_1$, its higher harmonics, and a zero harmonic $\boldsymbol{\Phi}_0$. The latter is also called a \textit{shift mode} in~\cite{noack2003hierarchy} as it represents the difference  between the mean state and the steady equilibrium. As $r(t)$ saturates to $r_\star$ and
$\Omega(t)\to\Omega_\star$, this representation reduces to the standard
time-periodic harmonic structure of the limit cycle, while away from
saturation it provides a nonlinear, slowly modulated extension of DMD that
remains valid throughout the transient.

\section{Results}\label{sec:Results}

We demonstrate the application of the proposed invariant-manifold-based reduced-order modeling approach to the
transonic buffet phenomenon over the OAT15A supercritical airfoil, a widely used benchmark configuration for
two-dimensional shock buffeting studies. High-fidelity unsteady Reynolds-averaged Navier--Stokes (URANS) simulations are employed to characterize buffet
onset and offset and to generate the datasets used for training and validation of the ROM, as detailed in
Section~\ref{sec:Dataset}. Using these data, we identify the invariant manifold and associated reduced dynamics in
Section~\ref{sec:ROMresults}. We also demonstrate modal decomposition here. Finally, Section~\ref{sec:StateReconstructionValidation} presents the full-state reconstruction of the flow and the validation of the reduced-order model.

\subsection{Buffet Characterization}\label{sec:Dataset}

URANS simulations are performed, assuming fully turbulent flow, using the DLR TAU code, an edge-based, unstructured finite-volume solver employing a cell-vertex scheme, as described in \cite{Schwamborn2006}.
Among the available turbulence closures, the negative Spalart–Allmaras one-equation turbulence model (SA–neg, \cite{Allmaras2012}) is employed.
Simulations are performed over a range of angles of attack from $4.5^\circ$ to $6^\circ$ to capture the onset and
offset of transonic buffet at a fixed freestream Mach number of $M_{\infty} = 0.71$ and Reynolds number of $Re=3 \times 10^6$ based on the chord length. 

\subsubsection{Numerical Settings, Spatial and Temporal Discretization}

The OAT15A supercritical airfoil, with chord length $c = 1~\mathrm{m}$, is modeled using an adiabatic no-slip wall boundary condition. The angle of attack is varied by adjusting the streamwise and transverse velocity components at the far-field boundary, prescribed as $(u,v) = U_\infty(\cos\alpha, \sin\alpha)$, where $U_\infty = 225.275~\mathrm{m/s}$ is the freestream reference velocity.

\begin{figure}[htbp]
\centering
\includegraphics[width=\textwidth,trim=17pt 580pt 50pt 20pt,clip]{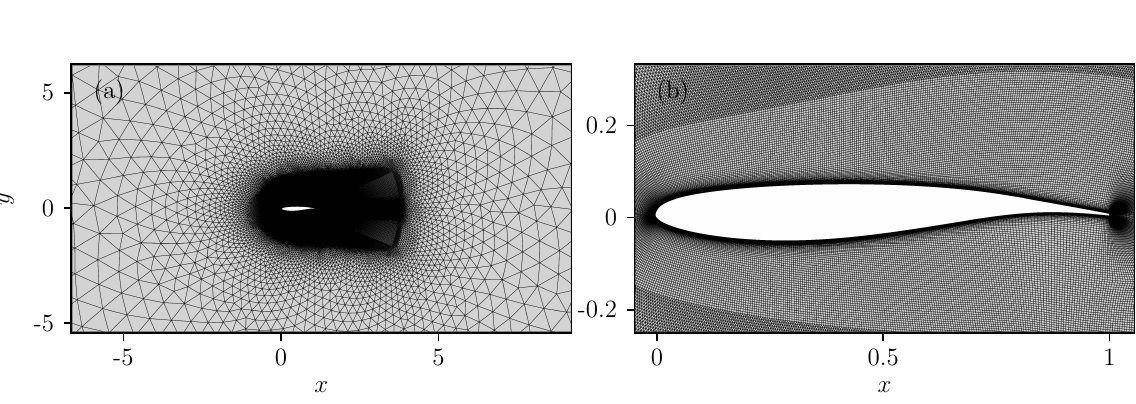}
\caption{Computational mesh employed in this study, adopted from \cite{nitzsche2019fluid}: (a) mesh in the vicinity of the airfoil and (b) close-up of the near-wall quasi-structured boundary-layer mesh.
}
\label{fig:mesh}
\end{figure}

The computational mesh employed in this study, shown in Figure~\ref{fig:mesh}, corresponds to the unstructured grid used in
\cite{nitzsche2019fluid}. 
The near-wall region consists of a body-fitted, quasi-structured O-type boundary-layer mesh, providing high resolution of the viscous sublayer with $y^+ < 1$. The quasi-structured region extends up to two chord lengths from the airfoil surface, maintaining an approximately isotropic spacing of $\Delta x \approx 0.5\%$ of the chord. Cell-stretching ratios within the boundary layer are limited to values below $1.2$ in all directions.
The far-field boundary is defined by a circular contour with a radius of $100c$, with the surrounding domain discretized using progressively coarsened unstructured triangular elements. The mesh comprises approximately $121{,}000$ nodes, corresponding to $n = 605{,}000$ degrees of freedom in the cell-vertex formulation, where control volumes are constructed around each grid point (\cite{Gerhold1997}). Inviscid convective fluxes are discretized using a second-order central scheme with artificial matrix dissipation (\cite{Jameson1981}), while the convective fluxes of the turbulence equation are computed using a first-order Roe upwind scheme. Spatial gradients are evaluated using a Green–Gauss formulation. Geometric multigrid acceleration is enabled through agglomerated coarsened grids generated within the DLR TAU code.

Time integration of the URANS equations is conducted using the dual-time-stepping approach from \cite{Jameson1991}. 
Physical time advancement is performed using a second-order backward differentiation formula (BDF2), with each physical time step formulated as a modified steady-state problem that is advanced in pseudo-time.
The pseudo-time iterations are efficiently advanced using local time stepping with an implicit backward Euler scheme and a lower-upper symmetric Gauss–Seidel (LU-SGS) solver.
Geometric multigrid acceleration is employed within the pseudo-time iterations to improve convergence rates.

The unsteady simulations are initialized from a converged steady-state base flow solution, obtained by iterating until the $l_2$-norm of the density residual reaches $10^{-10}$. 
Unsteady time integration is carried out using a constant physical time step $\Delta t=4.4 \cdot 10^{-4}$, selected based on a systematic time-step refinement study. A fixed number of inner pseudo-time iterations is applied at each physical time step to ensure consistent convergence behavior. Details regarding the choice of time step and convergence behavior are provided in \cite{jayaraj_2025_thesis}.

\subsubsection{Steady Base Flow}

\begin{figure}[htbp]
\centering
\includegraphics[width=0.5\textwidth, trim=100pt 510pt 100pt 50pt, clip]{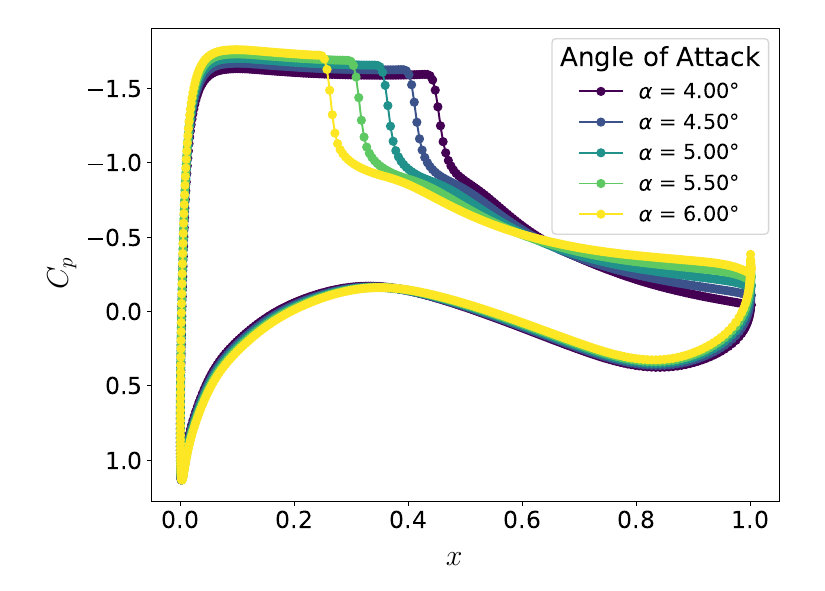}
\label{}
 \caption{Surface pressure coefficient $C_p$ distributions over the OAT15A airfoil for several angles of attack at $M_{\infty}=0.71$. The suction-side pressure plateau followed by a sharp gradient indicates the presence of a transonic shock, which moves upstream toward the leading edge with increasing angle of attack.}

         \label{fig:pressure}
\end{figure}

We begin by examining solutions of the steady RANS equations. Figure~\ref{fig:pressure} presents the surface pressure coefficient $C_p$ distributions for a range of angles of attack. A sharp gradient in $C_p$ on the suction side of the airfoil indicates the presence of a shock wave. Upstream of the shock, the pressure coefficient exhibits a nearly constant plateau, characteristic of a locally supersonic flow region terminated by the shock. With increasing angle of attack, the suction-side pressure level decreases slightly and the shock moves upstream toward the leading edge.

\begin{figure}[htbp]
\centering
 \includegraphics[width=\textwidth,trim=15pt 560pt 50pt 40pt,clip]{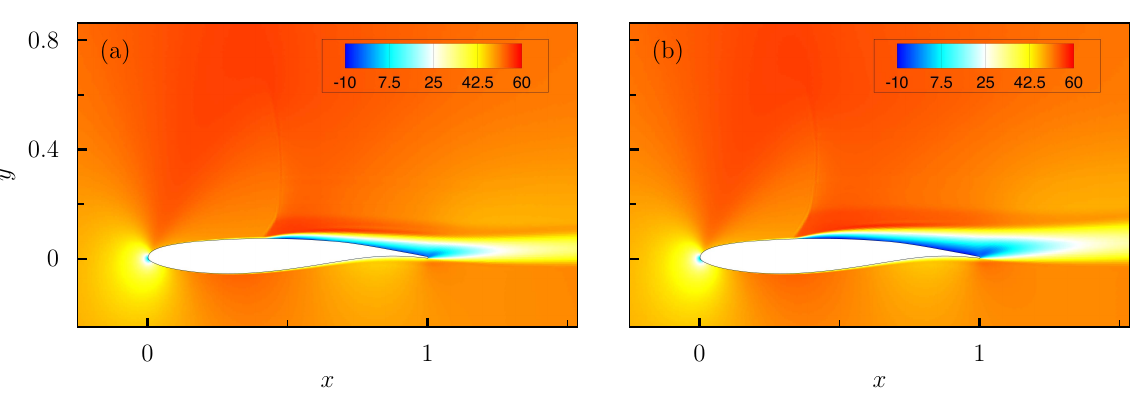}

\caption{Steady streamwise momentum $\rho u$ field at $M_{\infty}=0.71$ for (a) $\alpha = 4.45^\circ$ and (b) $\alpha = 5.25^\circ$. The sharp spatial gradient corresponds to the shock. Regions of reversed streamwise momentum downstream of the shock indicate flow separation. For the higher angle of attack shown, the shock is located further upstream on the suction side and the separated region has a larger spatial extent.}
 \label{fig:xmomentum_steady}
\end{figure}

Figure~\ref{fig:xmomentum_steady} shows the distributions of streamwise momentum $\rho u$ for the two angles of attack, $\alpha = 4.45^\circ$ and $\alpha = 5.25^\circ$. 
Regions of high momentum along the suction side correspond to the supersonic flow zone upstream of the shock, which appears as a sharp spatial gradient. Downstream of the shock, a separated flow region develops, characterized by  negative values of streamwise momentum. As the angle of attack increases and the shock shifts upstream, the separated region enlarges and extends further downstream.

\subsubsection{Unsteady Flow and Buffet Dynamics}

The URANS simulations can be initialized either from an unconverged steady solution or by applying a prescribed perturbation to the freestream, as arbitrary initialization of the unsteady flow state is not possible in the solver. The former approach would require prohibitively long computation times to reach the buffet limit cycle. We therefore adopt the latter approach and excite the flow using pitching-pulse perturbations, which are also representative of disturbances encountered in practical applications. The resulting simulations are validated against available results in the literature in~\cite{jayaraj_2025_thesis}.

\begin{figure}[htbp]
\centering
    \includegraphics[width=0.65\linewidth]{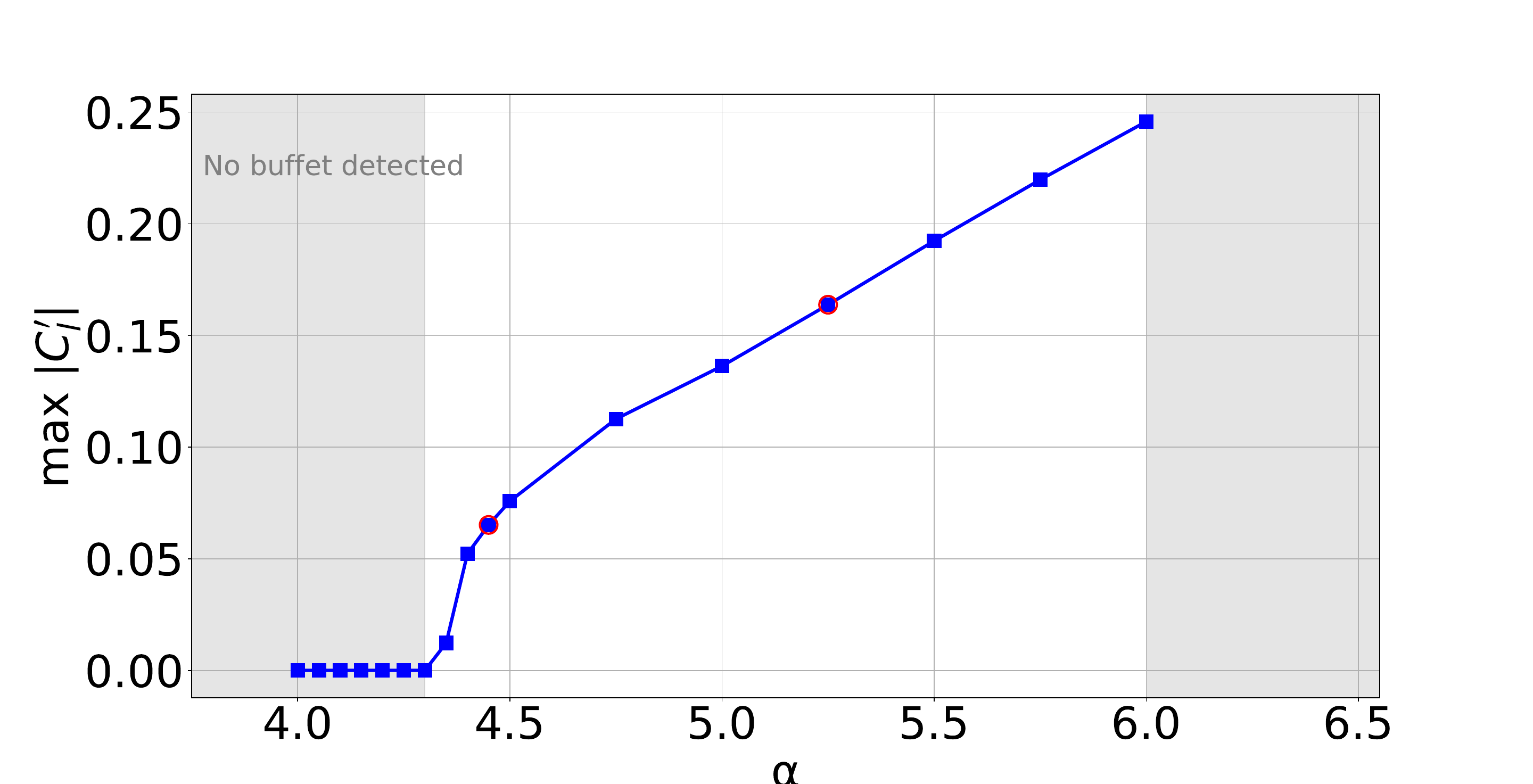}
   \caption{Bifurcation diagram of the maximum magnitude of the normalized lift coefficient, $\max_t |\bar{C_l}(t)| = \max_t \left| (C_l(t) - C_{l_s})/C_{l_s} \right|$, as a function of angle of attack $\alpha$ at $M_\infty = 0.71$. A smooth increase of $\max_t |\bar{C_l}(t)|$ from near-zero values is observed beginning at $\alpha \approx 4.35^\circ$, indicating the onset of transonic buffet. The two circled points denote the angles of attack selected for reduced-order model construction: $\alpha = 4.45^\circ$ (near buffet onset) and $\alpha = 5.25^\circ$ (mid-envelope).}
    \label{fig:buffet}
\end{figure}

Buffet onset and offset are identified by monitoring the maximum magnitude of the normalized unsteady lift coefficient,
\[
\bar{C_{l}}(t) = \frac{C_{l}'(t)}{C_{l_b}}, \qquad 
\max_t |\bar{C_{l}}(t)|,
\]
where $C_{l_b}$ denotes the corresponding steady lift coefficient. In the steady regime, $\max_t |\bar{C_{l}}(t)|$ remains close to zero, whereas it increases to a finite value once a self-sustained limit-cycle oscillation develops. Unsteady simulations are conducted over the range $\alpha \in [4^\circ, 6^\circ]$ in order to capture the buffet envelope. The buffet onset is identified at approximately $\alpha \approx 4.35^\circ$, where $\max_t |\bar{C_{l}}(t)|$ begins to increase smoothly from zero, indicating the emergence of a supercritical Hopf bifurcation, as shown in Figure~\ref{fig:buffet}. This observation is consistent with the transonic buffet boundary reported in \cite{nitzsche2022effect}. For identical initial conditions, the transient duration decreases and the oscillation amplitude increases with angle of attack up to approximately $\alpha = 5.5^\circ$, in agreement with the trends reported in Figure~\ref{fig:Nitzsche boundary}. Beyond this angle, the transient duration increases. Simulations are not extended beyond $\alpha = 6^\circ$, as buffet offset has already been reported in this regime (\cite{nitzsche2022effect}).

To enable independent training and validation, we generate two unsteady URANS trajectories at each of the two angles of attack selected for reduced-order model construction, $\alpha = 4.45^\circ$ (near buffet onset) and $\alpha = 5.25^\circ$ (mid-envelope). These are referred to as \textit{Trajectory~1} and \textit{Trajectory~2}, and are used for model training and validation, respectively. Distinct pulse amplitudes are used to produce independent trajectories.
The corresponding pulse amplitudes $\alpha_{\mathrm{pulse}}$ and the number of snapshots extracted from each simulation are summarized in Table~\ref{tab:pulse}. The training trajectories include approximately 8-10 limit-cycle oscillations. The duration of validation trajectories are truncated is chosen to be suitable for subsequent error analysis, as discussed later.

\begin{table}
  \begin{center}
\def~{\hphantom{0}}
  \begin{tabular}{cccc}
\hline
\textbf{Angle of attack} & \textbf{Trajectory} & 
\boldmath$\alpha_{\mathrm{pulse}}$ \textbf{[deg]} & 
\textbf{$N$} \\
\hline
$\alpha = 4.45^\circ$ & Trajectory 1 (training)   & 0.001 & 9422 \\
                      & Trajectory 2 (validation) & 0.010 & 7614 \\
$\alpha = 5.25^\circ$ & Trajectory 1 (training)   & 0.001 & 4843 \\
                      & Trajectory 2 (validation) & 0.010 & 4843 \\
\hline
\end{tabular}
  \caption{Pitching-pulse excitation amplitude $\alpha_{\mathrm{pulse}}$ and number of snapshots $N$ for the training (Trajectory~1) and validation (Trajectory~2) simulations at each angle of attack used for reduced-order model construction.}
  \label{tab:pulse}
  \end{center}
\end{table}

 Figures~\ref{fig:recon4.45} and~\ref{fig:recon5.25} present snapshots of the instantaneous streamwise momentum $\rho u$ extracted at three representative time instants along the validation trajectories for both angles of attack. These time instants correspond to characteristic phases of the buffet cycle (see Section~\ref{sec:NormalFormModes}). The unsteady momentum fluctuations are predominantly localized in the vicinity of the steady shock position and within the downstream separated region, indicating oscillatory shock motion coupled with the periodic growth and shrinkage of the separation zone.

\subsection{Reduced-Order Model}\label{sec:ROMresults}
We now apply the methodology from Section~\ref{sec:ROM} to construct an invariant-manifold-based ROM for transonic buffet. The results are organized in three parts.
In Section~\ref{sec:NormalFormModes}, the identified dynamics are transformed into an extended normal-form representation, enabling analytical insight into the dynamics and a physically interpretable modal decomposition (\textbf{Stage II}).

\subsubsection{Invariant Manifold and Dynamics Identification} \label{sec:ManifoldDynamicsIdentification}

\paragraph{Invariant Manifold}
The invariant manifold associated with the buffet dynamics is identified using Algorithm~\ref{alg:tangent_refinement} from a single training trajectory (Trajectory~1). A manifold polynomial order of $M=5$ is selected for both angles of attack, providing sufficient flexibility to capture nonlinear effects while maintaining numerical robustness. The initial tangent-plane approximation is chosen as the subspace spanned by the two leading POD modes associated with the limit-cycle oscillation. 
The tangent plane is recovered through iterative refinement, with convergence achieved within ten iterations for both operating conditions and the tangent-plane correction residual falling below $\epsilon = 10^{-8}$.

\begin{figure}[htbp]
   \centering
        \includegraphics[width=\textwidth,trim=50pt 370pt 20pt 60pt,clip]
     {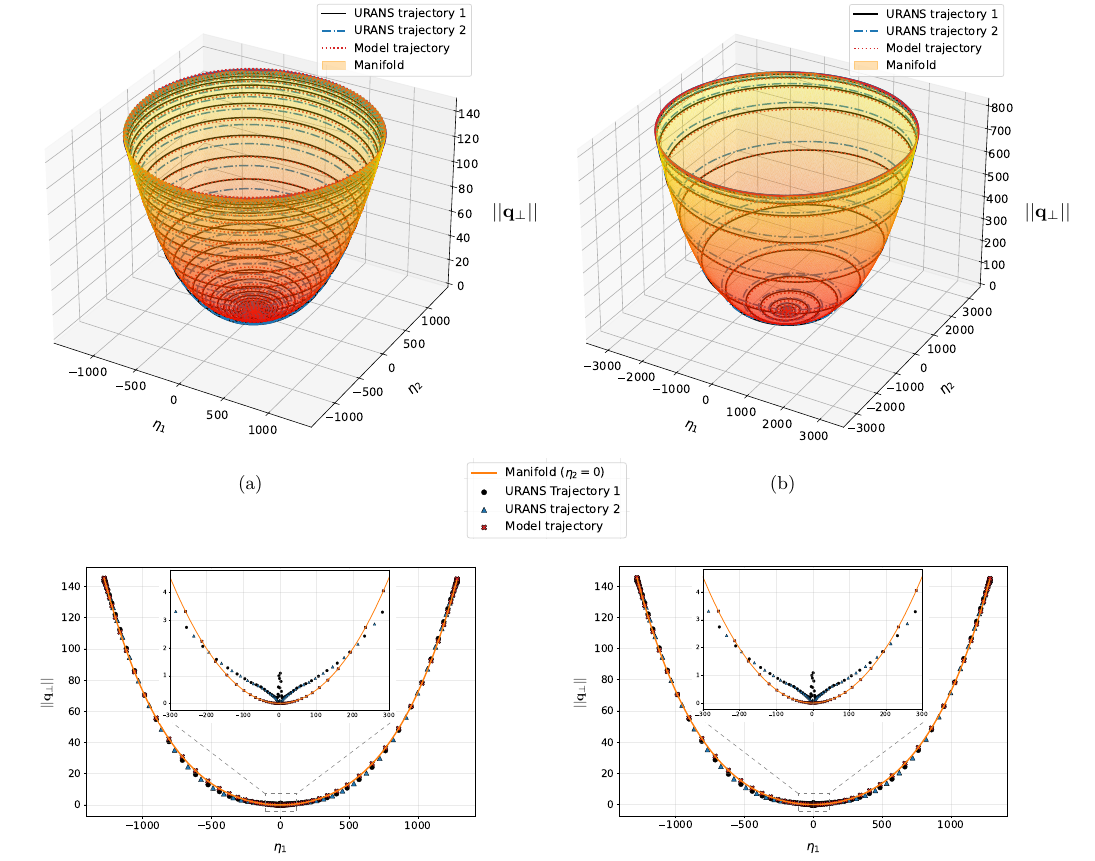}
        \includegraphics[width=\textwidth,trim=50pt 330pt 20pt 500pt,clip]
     {Plots/Manifold_plot.pdf}
     \vspace{-1em}
    \label{fig:manifold_xsec_4p45}
   \caption{Invariant manifold reconstruction for the two representative operating conditions, 
$\alpha = 4.45^{\circ}$ (left) and $\alpha = 5.25^{\circ}$ (right). 
Panels (a)–(b) show the reconstructed two-dimensional invariant manifold in graph–style coordinates 
$(\eta_{1},\eta_{2},\|\mathbf{x}_{\perp}\|)$, together with URANS trajectories and the corresponding ROM prediction. Trajectory 1 represents a training trajectory and Trajectory 2 the validation trajectory. 
Panels (c)–(d) present cross-sections of the manifold at $\eta_{2}=0$ with a zoomed region near the equilibrium, illustrating the convergence of pulse-initialized trajectories onto the manifold. 
}
    \label{fig:Manifolds}
\end{figure}

The identified manifolds are shown in Figure~\ref{fig:Manifolds}(a)–(b), plotted in graph-style reduced coordinates $(\eta_1,\eta_2)$ with the vertical axis representing the norm of the component orthogonal to the estimated tangent space,
\[
\|\mathbf{q}_{\perp}\| = \|\mathbf{q} - \mathbf{W}_1 \mathbf{W}_1^T \mathbf{q}\|,
\]
thereby illustrating the deviation of the manifold from its tangent plane. For both angles of attack, the corresponding URANS training and validation trajectories approximately lie on the recovered manifold, confirming the existence of a smooth two-dimensional invariant manifold connecting the steady equilibrium and the limit-cycle attractor, even far from the bifurcation point. Furthermore, the manifolds do not fold over the approximated tangent plane. 
At the higher angle of attack, the surface shows larger deviation from the initial tangent plane, with the orthogonal component $\|\mathbf{q}_{\perp}\|$ reaching approximately six times the magnitude of the lower-$\alpha$ case. 
Near the equilibrium, the trajectories initially undergo a pulse-induced transient before converging onto the manifold surface, as visible from the manifold cross-sections in Figure~\ref{fig:Manifolds}(c)–(d). 
The largest reconstruction discrepancies are observed near the base-flow state, arising from two factors: 
(i) the early transient phase, during which the trajectory has not yet settled onto the true attracting manifold, and 
(ii) the intrinsic difficulty of estimating a high-dimensional manifold from a single trajectory that does not originate close to the equilibrium. 
As the trajectory approaches the limit cycle, the approximation accuracy improves significantly.
The lower-angle case exhibits smaller discrepancies overall, as expected for a weaker nonlinear regime. A quantitative evaluation of the modeling accuracy is provided later in Section~\ref{sec:StateReconstructionValidation}.

\begin{figure}
\centering
    \includegraphics[width=\textwidth,trim=20pt 50pt 20pt 21pt,clip]{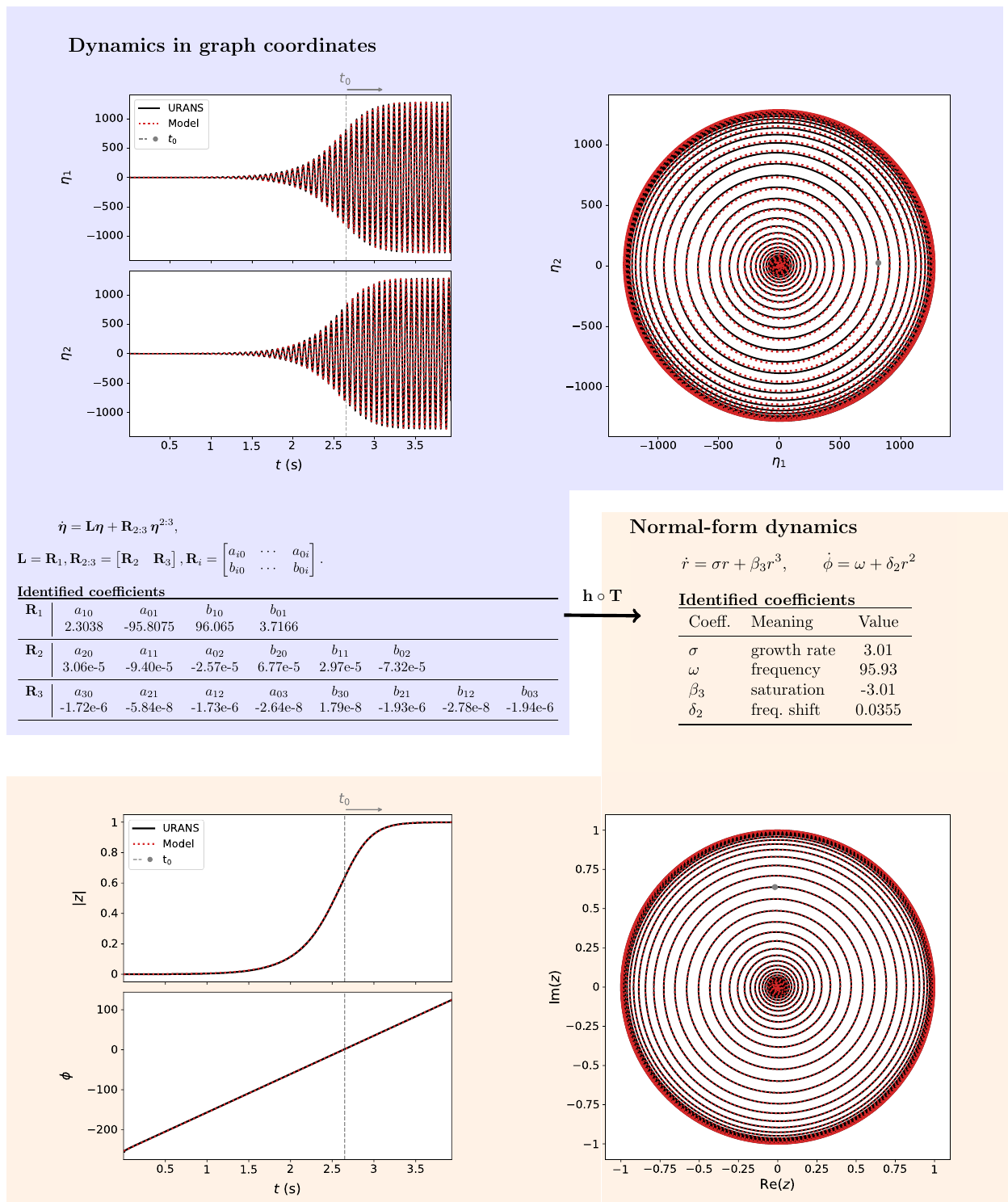}
\caption{Identification of reduced dynamics and normal-form representation for transonic buffet at
 $\alpha = 4.45^\circ$.
Top: Identified reduced-order model in graph-style coordinates $\boldsymbol{\eta}$ compared with URANS training trajectory $\boldsymbol{\eta}_{\mathrm{URANS}}(t)$ projected onto the estimated tangent plane. Middle: Coefficients of the identified reduced-order model in graph-style coordinates (left) and normal-form coefficients obtained via the transformation $h \circ T$ (right). Bottom: Amplitude–phase (Stuart--Landau) dynamics of the reduced system in normal-form coordinates compared with
URANS.
}
\label{fig:polar_validation_w_coeff_tables}
\end{figure}

With the invariant manifold identified, the reduced dynamics on the manifold are inferred from the projected URANS training trajectories $\boldsymbol{\eta}_{\mathrm{URANS}}(t)$ (Trajectories~1) by fitting the polynomial reduced-order model described in Section~\ref{sec:ROM}. To improve estimation accuracy, only the portion of each trajectory starting at time $t_0$ is retained, where pulse-induced transients have decayed and the manifold approximation is most reliable. For both angles of attack, a third-order polynomial expansion accurately represents the reduced dynamics. The resulting coefficients for the $\alpha = 4.45^\circ$ case are listed in Figure~\ref{fig:polar_validation_w_coeff_tables} (middle row, left table), while the coefficients for the other operating condition are provided in the Supplementary Material. Using these coefficients, projected Trajectory~1 is reconstructed by forward and backward time integration of the identified ODE, initialized at the state corresponding to $t_0$.
As shown in Figure~\ref{fig:polar_validation_w_coeff_tables} (top row), it exhibits excellent agreement with the URANS data. As further
illustrated in Figure~\ref{fig:Manifolds}, the reconstructed Trajectory~1,
embedded back into the full state space, closely matches the corresponding
URANS trajectory, confirming the fidelity of the identified reduced-order model.
Only the training trajectory is shown here, as it is used to identify the
reduced dynamics; the validation trajectory is examined in
Section~\ref{sec:NormalFormModes} once the extended normal-form transformation is
introduced.

\subsubsection{Reduced Dynamics in Normal-Form and Modal Decomposition}
\label{sec:NormalFormModes}

After identifying the reduced dynamics in graph-style coordinates $\boldsymbol{\eta}$,
we apply extended normal-form transform explained in Section~\ref{sec:extendednormalform}. 
To improve numerical conditioning, the columns of the linear transformation
matrix \(\mathbf{T}\) are scaled such that the limit-cycle amplitude in the
transformed coordinates satisfies \(r_{\star} = 1\). The transformation is
truncated at third order, yielding the Stuart--Landau equations. The identified
coefficients are reported in
Figure~\ref{fig:polar_validation_w_coeff_tables} (middle row, right table) for $\alpha=4.45^{\circ}$ (for $\alpha=5.25^{\circ}$ see Supplementary Materials).

To assess the predictive accuracy of the identified dynamics, the solutions of
the Stuart--Landau equations are compared against the projected  validation URANS
trajectory (Trajectory~2). The Stuart--Landau model is evaluated forward and backward in time from two representative initial conditions selected along the validation trajectory: \(t_0'\), located in the early transient phase following the decay of the pulse-induced response, and \(t_0\), corresponding to a later transient state closer to the limit-cycle.
For direct comparison, the projected validation URANS trajectory is mapped into the
normal-form coordinates by applying the inverse transformation to the reduced
data, i.e.,
$\mathbf{z}_{\mathrm{URANS}}(t)
= \mathbf{T}^{-1}\!\circ\mathbf{h}^{-1}
\bigl(\boldsymbol{\eta}_{\mathrm{URANS}}(t)\bigr)$.

\begin{figure}[htbp]
\centering
\begin{subfigure}[b]{0.8\textwidth}
    \centering
    \includegraphics[width=\textwidth,trim=10pt 0pt 0pt 0pt,clip]   {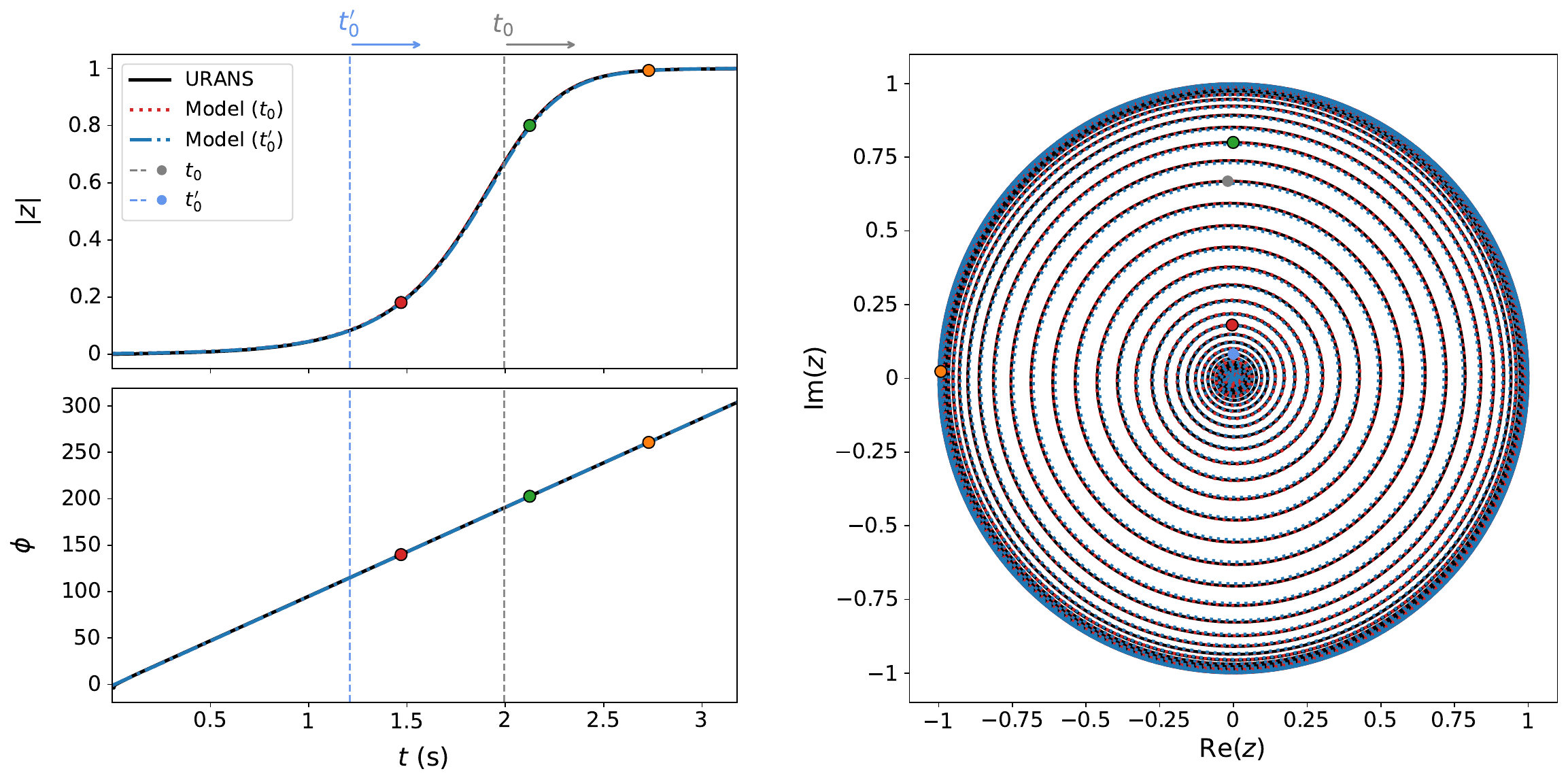}
    \caption{Validation of reduced dynamics at $\alpha = 4.45^\circ$}
    \label{fig:polar_validation_445}
\end{subfigure}

\vspace{0.45cm}

\begin{subfigure}[b]{0.8\textwidth}
    \centering
    \includegraphics[width=\textwidth,trim=10pt 10pt 0pt 0pt,clip] {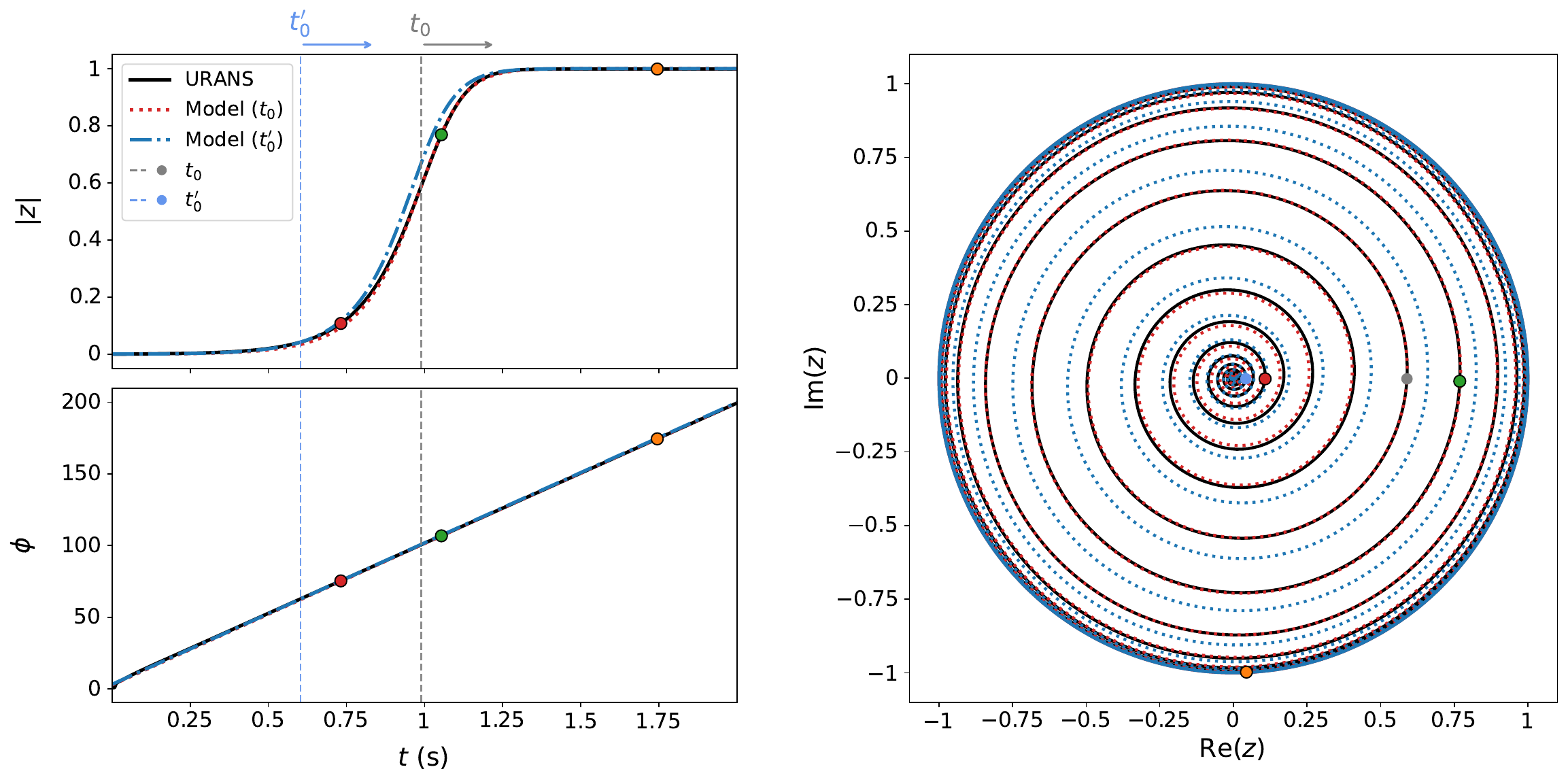}
    \caption{Validation of reduced dynamics at $\alpha = 5.25^\circ$}
    \label{fig:polar_validation_525}
\end{subfigure}

\caption{
Validation of reduced dynamics in extended normal-form (Stuart-Landau model).
\textbf{(a)} $\alpha = 4.45^\circ$, \textbf{(b)} $\alpha = 5.25^\circ$.
Model trajectories (red, blue) and URANS trajectories (black) are compared starting from multiple initial conditions. 
Markers indicate the three representative time instants used later for full-state reconstruction: 
early transient, approaching the limit cycle, and saturated oscillation.
}
\label{fig:polar_validation}
\end{figure}

Figure~\ref{fig:polar_validation} shows close agreement between the Stuart--Landau prediction and the URANS reference, both throughout the transient and at the saturated oscillation when the ROM is initialized at $t_0$. For the lower angle of attack this accuracy is also maintained when initializing closer to the equilibrium, whereas for the higher angle of attack the transient predicted from $t_0'$ shows slightly reduced accuracy.

To quantify the discrepancy between two trajectories
$\mathbf{y}_1(t)$ and $\mathbf{y}_2(t)$, we introduce the
\emph{mean weighted trajectory error} (MWTE):
\begin{equation}
\mathrm{MWTE}(\mathbf{y}_1,\mathbf{y}_2)
=
\frac{1}{N} \sum_{k=1}^{N}
\frac{\left\| \mathbf{y}_1(t_k)-\mathbf{y}_2(t_k)\right\|}
     {\left\| \mathbf{y}_2(t_k)\right\|},
\label{eq:WMTE}
\end{equation}
where normalization by $\|\mathbf{y}_2(t_k)\|$ acts as a time-dependent weight. 
MWTE is closely related to the NMTE of~\cite{cenedese2022data},
but uses this normalization to avoid understating errors near the equilibrium (where the state magnitude is small) relative to those near the limit cycle. 
For validation of the Stuart--Landau model, we set
$\mathbf{y}_1(t)=\mathbf{z}_{\mathrm{ROM}}(t)$ and
$\mathbf{y}_2(t)=\mathbf{z}_{\mathrm{URANS}}(t)$.
The validation trajectories are truncated such that the duration of the
transient phase is approximately equal to the time spent on the limit cycle.
The resulting MWTE values, reported in
Table~\ref{tab:NFerrorvalues}, indicate small reconstruction errors for both angles of attack.

\begin{table}
  \begin{center}
\def~{\hphantom{0}}
  \begin{tabular}{cccc}
\hline
\textbf{Angle of attack} & \textbf{Projected ROM trajectory} & 
$\mathrm{MWTE}_\mathrm{fwd}$ & 
$\mathrm{MWTE}_\mathrm{bwd}$\\
\hline
\multirow{2}{*}{$\alpha = 4.45^\circ$}
  & ROM from $t_0'$ (early transient) & 2.89e-06 & 1.21e-06 \\
  & ROM from $t_0$ (late transient)   & 1.53e-06  & 1.36e-07  \\
\multirow{2}{*}{$\alpha = 5.25^\circ$}
  & ROM from $t_0'$ (early transient) & 6.34e-05 & 3.1e-06 \\
  & ROM from $t_0$ (late transient)   & 1.95e-05  & 3.1e-07\\
\hline
\end{tabular}
  \caption{Mean weighted trajectory error (MWTE) between the normal-form ROM
and the validation URANS trajectory (Trajectory~2), computed using
\eqref{eq:WMTE}. For both angles of attack, the ROM is initialized
in the early transient near the equilibrium ($t_0'$) and in the late
transient near the limit cycle ($t_0$), and solved forward and
backward in time.}
  \label{tab:NFerrorvalues}
  \end{center}
\end{table}

\paragraph{Modal interpretation.}

In normal-form coordinates, the manifold expansion yields a harmonic decomposition of the flow into a shift mode, a fundamental oscillatory mode, and higher-order harmonics associated with the buffet frequency. Under the assumption of slowly varying amplitude (Section~\ref{HopfInterpretation}), this decomposition is valid from the fixed point to the limit cycle. 

For the lower angle of attack (\(\alpha = 4.45^\circ\)), which lies close to the Hopf bifurcation, this assumption is satisfied during the transient, as the oscillation amplitude varies slowly over a single period (Figure~\ref{fig:polar_validation_445}). In contrast, for the higher angle of attack (\(\alpha = 5.25^\circ\)), the amplitude evolves more rapidly during the transient stage (Figure~\ref{fig:polar_validation_525}), violating the slow-variation assumption and restricting the modal interpretation to the vicinity of the equilibrium and the saturated limit-cycle regime.

Accordingly, we present the dominant eigenmodes near equilibrium and the harmonic modes on the limit cycle for both angles of attack, and additionally include harmonic modes at a representative transient state for \(\alpha = 4.45^\circ\).

\begin{figure}[t]
          \centering
         \includegraphics[width=\textwidth, trim=0pt 450pt 0pt 0pt]{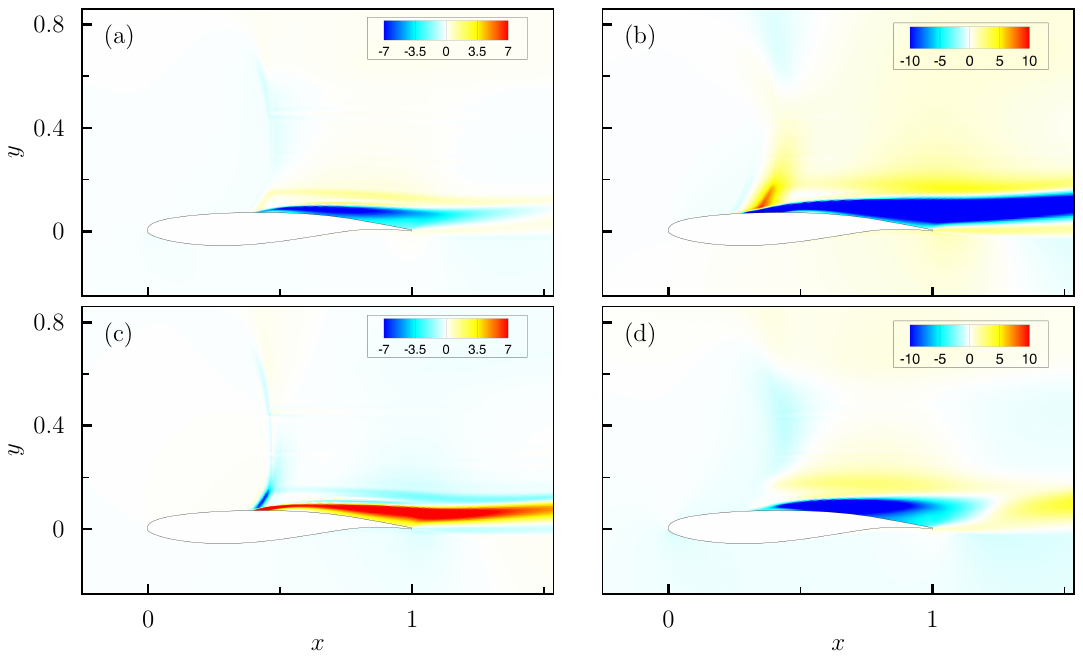}
         \caption{
Spatial structure of the principal buffet eigenmode $\tilde{\mathbf{W}}_1$, 
showing the streamwise momentum component. 
The real part (top panels) and imaginary part (bottom panels) are plotted for two 
operating conditions: (a,c) $\alpha = 4.45^{\circ}$ and (b,d) $\alpha = 5.25^{\circ}$.}\label{fig:eigenmodes}
\end{figure}

The complex-conjugate eigenvector pair $\tilde{\mathbf{W}}_1$, shown in Figure~\ref{fig:eigenmodes} for both angles of attack, captures the spatial structure of the buffet instability in the streamwise momentum field near the base flow. The mode is concentrated in the shock region and the downstream recirculation zone. Notably, the streamwise momentum perturbations exhibit opposite signs in these regions, indicating synchronized oscillations in which downstream shock motion is accompanied by contraction of the separation region, and vice versa.

\begin{figure}[t]
          \centering
         \includegraphics[width=\textwidth, trim=0pt 380pt 0pt 0pt]{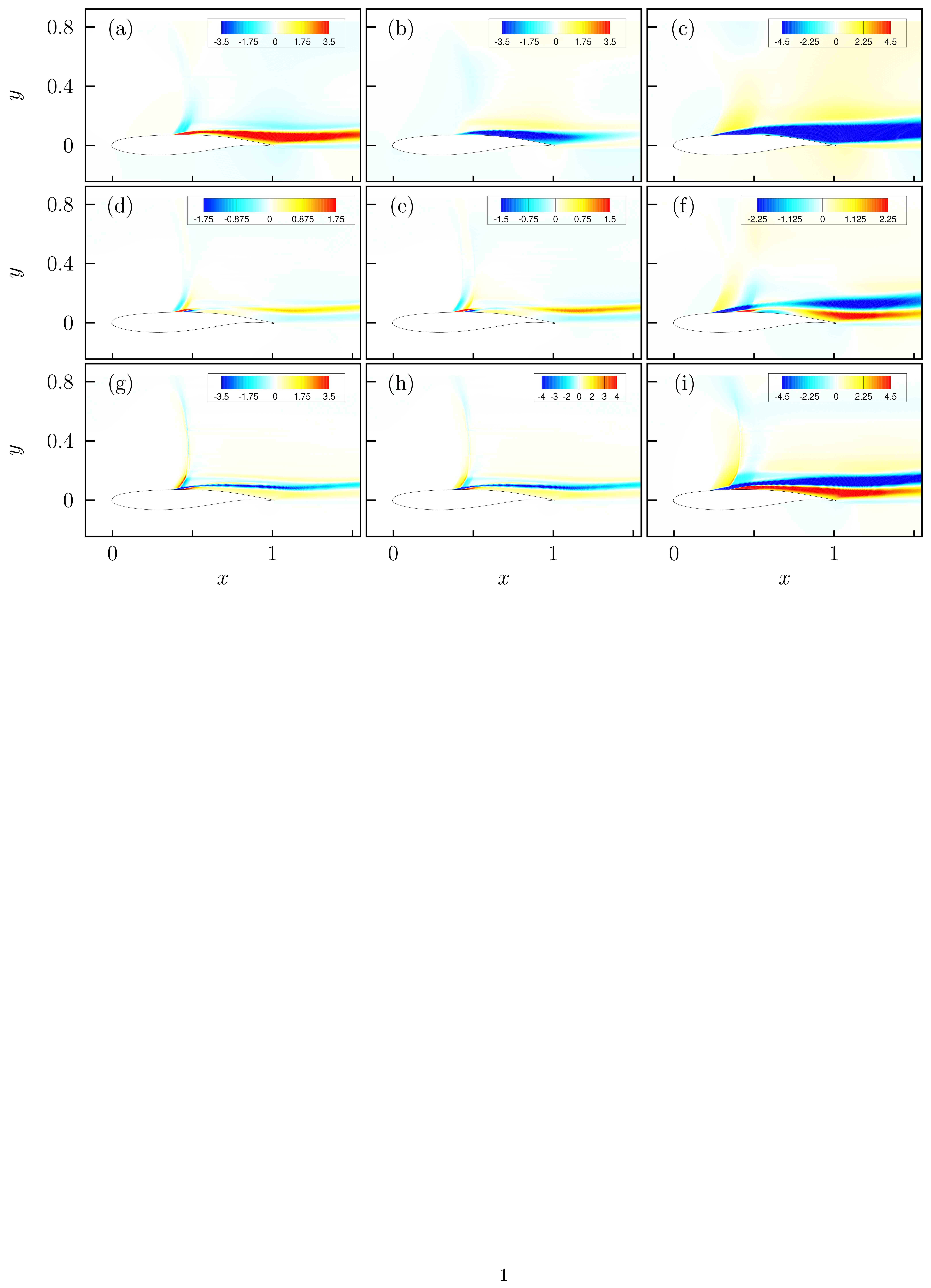}
         \caption{
Spatial structure of the modal decomposition, 
showing the streamwise momentum component. 
Panels show the fundamental mode $\boldsymbol{\Phi}_1$ ((a)–(c)), second harmonic $\boldsymbol{\Phi}_2$ ((d)–(f)), and shift mode $\boldsymbol{\Phi}_0$ ((g)–(i)). The first column ((a), (d), (g)) corresponds to a transient state at $\alpha=4.45^{\circ}$. The second and third columns show the same modes on the saturated limit cycle at $\alpha=4.45^{\circ}$ and $\alpha=5.25^{\circ}$, respectively.}\label{fig:harmonics}
\end{figure}

Figure~\ref{fig:harmonics} (second and third column) shows the harmonic modes $\boldsymbol{\Phi}_k$ obtained from the limit-cycle evaluation of the manifold expansion. The fundamental harmonic $\boldsymbol{\Phi}_1$ corresponds to the primary buffet oscillation and captures the dominant oscillatory structure of the limit cycle. At the lower angle of attack, $\boldsymbol{\Phi}_1$ closely resembles the eigenmode, whereas at the higher angle of attack it exhibits more pronounced deformation, reflecting the increasing influence of nonlinear saturation. Compared to the eigenmode, $\boldsymbol{\Phi}_1$ is active over a wider region around the shock as observed in the nonlinear regime and previously reported in~\cite{sansica2022system}. The second harmonic $\boldsymbol{\Phi}_2$ and the mean (shift) mode $\boldsymbol{\Phi}_0$ exhibit finer spatial scales than $\boldsymbol{\Phi}_1$, consistent with their higher harmonic content. For \(\alpha = 4.45^\circ\), similar behavior is observed at a representative transient state shown in Figure~\ref{fig:harmonics} (first column). 
For space considerations, the harmonic modes are visualized using either their real or imaginary parts, chosen based on the reconstruction time instants (shown in Figure~\ref{fig:polar_validation}) used later for full-state analysis to facilitate interpretation of their spatial contributions.

The fundamental limit-cycle frequency predicted by the normal-form is compared with the dominant spectral peak obtained from an Fast Fourier Transform (FFT) of the URANS lift coefficient (Table~\ref{tab:frequency}) and shows good agreement. The corresponding Strouhal number is $St \approx 0.06$, consistent with values reported for this airfoil in~\cite{deck2005numerical}, \cite{jacquin2009experimental}, \cite{fukushima2018wall}, and \cite{cuong2022large}.

\begin{table}
  \begin{center}
\def~{\hphantom{0}}
  \begin{tabular}{ccc}
\hline
\textbf{Angle of attack} & \textbf{ ROM frequency} $\Omega_\star$ & 
URANS frequency $\Omega_{\mathrm{URANS}}$  \\
\hline
$4.45^\circ$ & 95.9692 & 95.5517 \\
$5.25^\circ$ & 97.8804 & 97.9571 \\
\hline
\end{tabular}
  \caption{Comparison of limit-cycle frequencies obtained from the normal-form ROM 
and from the dominant spectral peak of the URANS lift coefficient.}
  \label{tab:frequency}
  \end{center}
\end{table}

\subsection{Full–State Reconstruction and ROM Validation}
\label{sec:StateReconstructionValidation}
Here we demonstrate the predictive capabilities of the ROM through reconstruction of the full high-dimensional flow state.
This is assessed by comparing the reconstructed high-dimensional flow field,
\[
\mathbf{q}_{\mathrm{ROM}}(t)=\tilde{\mathbf{W}}\big(\mathbf{z}(t)\big),
\]
with the reference URANS solution (Trajectory~2). The reconstruction accuracy is evaluated using the weighted trajectory error (WTE), the instantaneous counterpart of MWTE~\eqref{eq:WMTE}
\begin{equation}
\mathrm{WTE}(t_k)
    = 
    \frac{\| y_1(t_k)-y_2(t_k)\|}{\|y_2(t_k)\|},
\end{equation}
now applied to the full-state trajectories by setting
$\mathbf{y}_1(t)=\mathbf{q}_{\mathrm{ROM}}(t)$ and
$\mathbf{y}_2(t)=\mathbf{q}_{\mathrm{URANS}}(t)$.

\begin{figure}[t]
         \centering
         \includegraphics[width=\textwidth, trim=80pt 590pt 100pt 80pt]{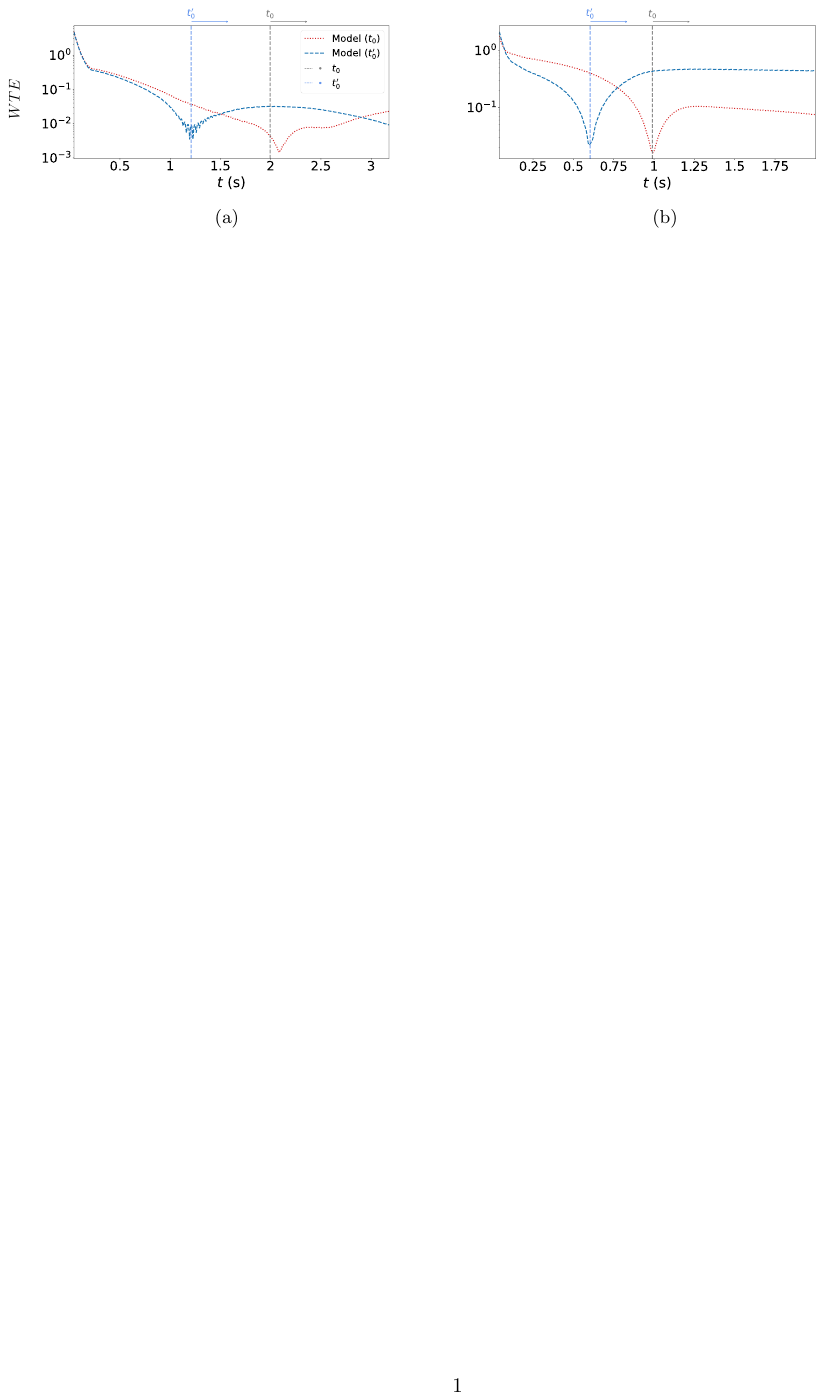}
         \caption{Instantaneous weighted trajectory error (WTE), 
$\|y_1(t_k)-y_2(t_k)\|_2 / \|y_2(t_k)\|_2$, for two representative initial 
conditions $t_0$ and $t_0'$ indicated with vertical dashed lines.Panel (a) corresponds to $\alpha=4.45 ^{\circ}$ 
 and panel (b) to $\alpha=5.25 ^{\circ}$ 
. The error rapidly decays during attraction to the invariant 
manifold and varies slowly near the limit cycle. }

         \label{fig:errors}
\end{figure}

The resulting error values, evaluated using both $t_0'$ and a later initialization time $t_0$ from Figure~\ref{fig:polar_validation}, are shown in Figure~\ref{fig:errors}. For both angles of attack, the error is largest in the immediate vicinity of the base flow. This is expected, as the pulse excitation drives the flow away from the invariant manifold, and the resulting pulse-initialized trajectory therefore does not lie on the approximated manifold in this region, as observed earlier in Figure~\ref{fig:Manifolds}.

For \(\alpha = 4.45^\circ\), the error decreases rapidly during the early transient and, beyond \(t_0'\) (corresponding to \(r \approx 0.077\), or about $7.7\%$ of the limit-cycle amplitude), remains below $5\%$ for both initialization times. This indicates that, at this angle of attack, the ROM reconstructs the flow field with high accuracy from early nonlinear transients through to the limit cycle. 
For \(\alpha = 5.25^\circ\), the reconstruction errors are higher. When initialized at the early transient stage \(t_0'\), forward-time integration yields errors in the range of 20–50\%. However, when initialized later in the transient at $t_0$, at approximately 60\% of the limit-cycle amplitude, the reconstruction error is significantly reduced and remains below 10\%. This indicates that, for this angle of attack, the ROM provides accurate flow reconstruction from the mid-to-late transient regime and on the limit cycle.

\begin{figure}[t]
         \centering
         \includegraphics[width=0.8\textwidth, trim=50pt 330pt 50pt 0pt]{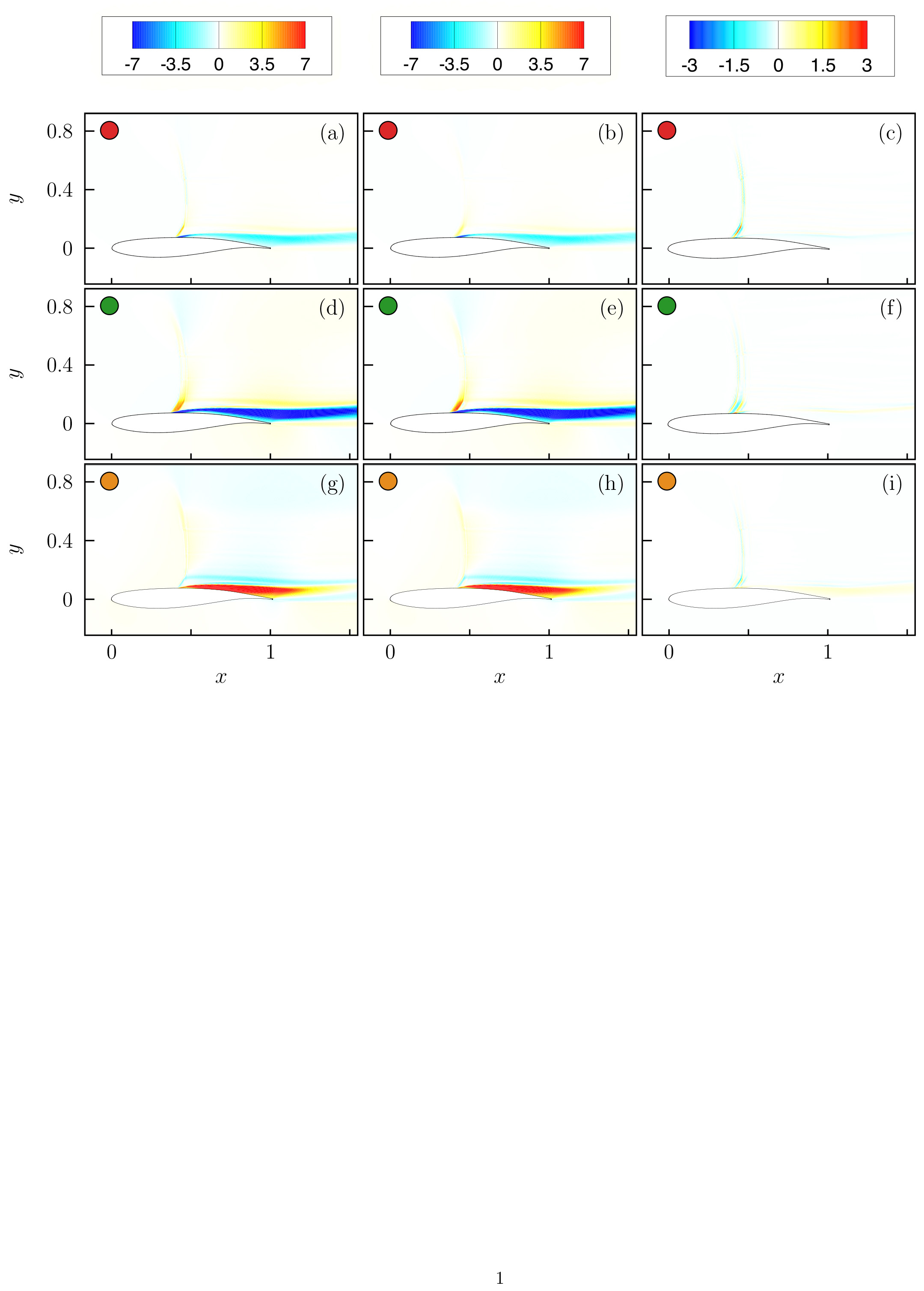}
         \caption{Reconstruction of the streamwise momentum field for $\alpha=4.45^{\circ}$ at selected points.  
Each column displays: (left) the URANS snapshot, (middle) the corresponding ROM 
reconstruction, and (right) the pointwise difference between the two fields.  
Each row corresponds to a specific time instant indicated by the colored dots 
(red, green, orange) in the time-evolution Figure~\ref{fig:polar_validation_445}.}
         \label{fig:recon4.45}
\end{figure}

\begin{figure}[t]
         \centering
         \includegraphics[width=0.8\textwidth, trim=50pt 350pt 50pt 0pt]{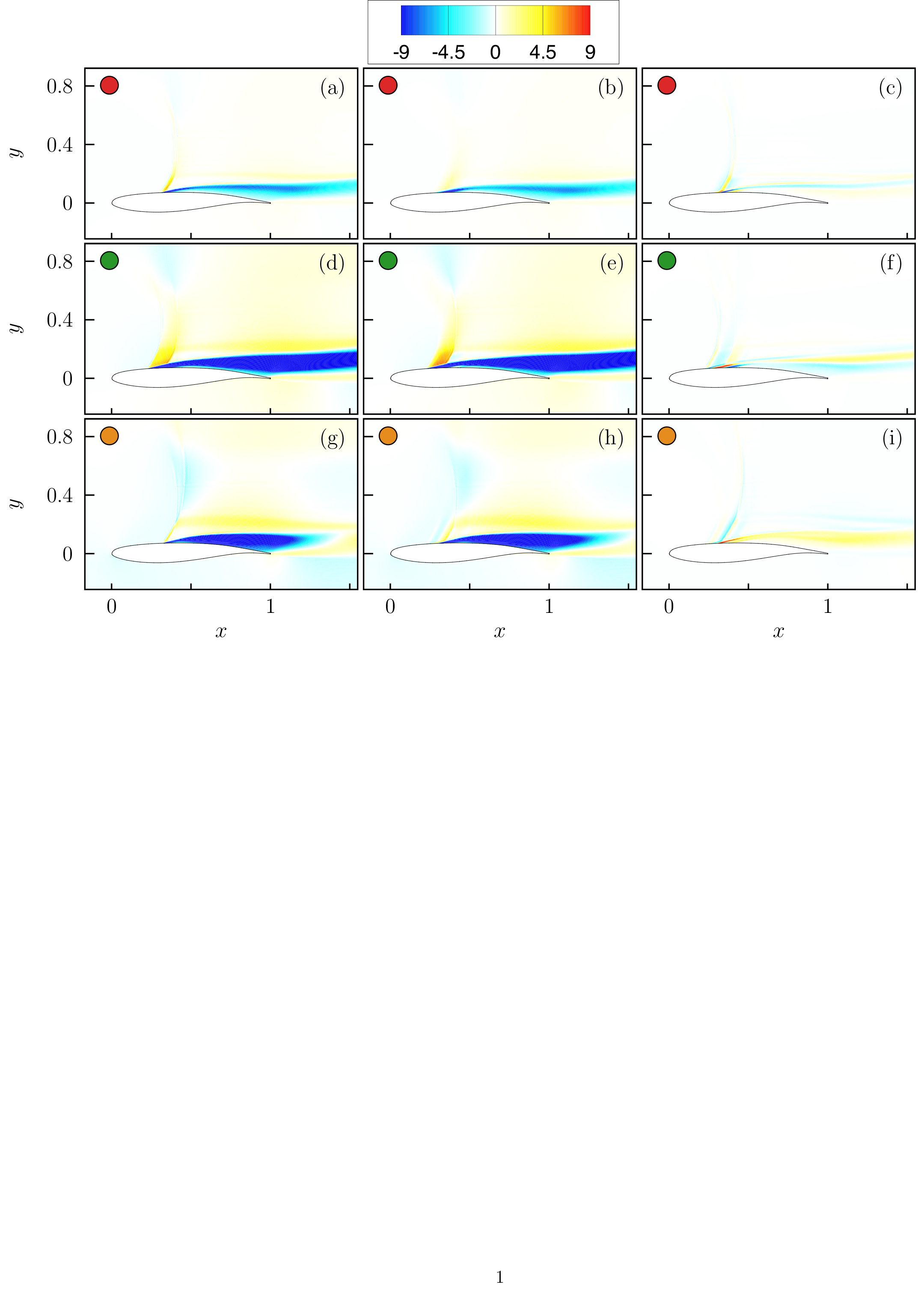}
         \caption{Reconstruction of the streamwise momentum field for $\alpha=5.25^{\circ}$ at selected points.  
Each column displays: (left) the URANS snapshot, (middle) the corresponding ROM 
reconstruction, and (right) the pointwise difference between the two fields.  
Each row corresponds to a specific time instant indicated by the colored dots 
(red, green, orange) in the time-evolution Figure~\ref{fig:polar_validation_525}.}
         \label{fig:recon5.25}
\end{figure}

Figures~\ref{fig:recon4.45} and~\ref{fig:recon5.25} illustrate the reconstruction of the streamwise momentum field \((\rho u)\) at three representative time instants (introduced in Figure~\ref{fig:polar_validation}). These instants are chosen such that either the real or imaginary component of the dominant modes, shown in the Figure~\ref{fig:harmonics}, is active, facilitating a clearer interpretation of their spatial contributions. The large-scale spatial patterns in the flow resemble those of the fundamental mode. Moreover, the overall reconstruction quality is high, with only small differences between the URANS and ROM fields, particularly for the late-transient and limit-cycle time instants.

\section{Conclusion} \label{sec:Conclusion}

In this work, we developed a reduced-order model (ROM) for two-dimensional transonic buffet. The aim was to address limitations of existing models by constructing a physically interpretable ROM that captures the nonlinear dynamics and enables reconstruction of the full flow state across the dynamically relevant region of phase space, from the unstable equilibrium to the saturated limit cycle.

This was achieved by exploiting the existence of a two-dimensional attracting invariant manifold. To this end, we employed a data-driven framework for identifying invariant manifolds and the reduced dynamics evolving on them from full flow-state measurements. This constitutes an adaptation of the framework introduced in~\cite{cenedese2022data}, tailored to large-scale CFD applications. This includes settings in which obtaining trajectories sufficiently close to the unstable equilibrium or densely sampling the relevant region of phase space would incur prohibitive computational cost, and where prior knowledge of suitable reduced coordinates is unavailable. The invariant manifold was identified as a graph over its tangent space through an iterative update of the linear encoder, with each iteration involving only a low-dimensional least-squares problem independent of the full state dimension. The reduced dynamics evolving on the manifold were identified from projected trajectory data and subsequently transformed into extended normal-form coordinates, yielding a Stuart--Landau model and enabling a physically interpretable nonlinear modal decomposition.

The test case considered is transonic buffet over the OAT15A supercritical airfoil at a fixed Mach number of $M_{\infty}=0.71$. The flow is simulated using URANS with the  DLR TAU code. The reduced-order model was identified for two angles of attack, using a single training trajectory for each operating condition.
 Validation against independent URANS trajectories showed close agreement in reduced coordinates, with weighted mean trajectory errors of order $10^{-7}$–$10^{-5}$. Full-state reconstruction errors were below $5\%$ for $\alpha = 4.45^{\circ}$ from early nonlinear transients onward, and below $10\%$ for $\alpha = 5.25^{\circ}$ when initialized in the mid-to-late transient regime. The reduced dynamics expressed in Stuart--Landau normal-form accurately reproduced the buffet frequency and enabled a physically interpretable decomposition of the flow into a shift mode, a fundamental oscillatory mode, and higher harmonics. For the lower angle of attack, which lies close to the buffet onset, this modal decomposition remained meaningful throughout the transient evolution, as the amplitude variable of the Stuart--Landau dynamics evolved slowly over a single oscillation period. For the higher angle of attack, corresponding to a mid-buffet regime at the given Mach number, the Stuart--Landau amplitude varied more rapidly during the transient evoluation, limiting the modal interpretation to the vicinity of the equilibrium and the saturated limit-cycle regime.

Overall, these results show that invariant-manifold-based reduced-order models can provide accurate, physically interpretable, and computationally efficient representations of transonic buffet dynamics. Future work will focus on extending the approach to parametric modeling across operating conditions and exploring its integration into control-oriented applications.

\vspace{\baselineskip}
\noindent\textbf{Declaration of interests.}
The authors report no conflicts of interest.

\newpage
\phantom{.}
\newpage
\appendix
\section{Efficient Computation of the Least-Squares Update}\label{appA}

Consider the convex optimization problem
\begin{equation*}
\min_{\mathbf{W}_1,\,\mathbf{W}_{2:M}}\sum_{k=1}^{N}
\big\|
\mathbf{q}_k
-
\mathbf{W}_1\,\boldsymbol{\eta}_k
-
\mathbf{W}_{2:M}\boldsymbol{\eta}_k^{2:M}
\big\|^2 .
\label{eq:appendix_main}
\end{equation*}
Using the notation
\begin{equation*}
\widehat{\boldsymbol\eta}_k
=
\begin{bmatrix}
\boldsymbol{\eta}_k^{\top} &
(\boldsymbol{\eta}_k^{2:M})^{\top}
\end{bmatrix}^{\top}
\in \mathbb{R}^{K} \qquad
\mathbf{W}
=
\begin{bmatrix}
\mathbf{W}_1 & \mathbf{W}_{2:M}
\end{bmatrix}
\in \mathbb{R}^{n \times K}.
\end{equation*}
the problem is cast in the compact form
\begin{equation}
\min_{\mathbf{W}}
\sum_{k=1}^{N}
\|
\mathbf{q}_k - \mathbf{W}\widehat{\boldsymbol\eta}_k
\|^2 .
\label{eq:appendix_ls}
\end{equation}
Exploiting the identity
\[
\mathbf{W}\widehat{\boldsymbol\eta}
=
(\widehat{\boldsymbol\eta}^{\top} \otimes \mathbf{I}_n)\,
\mathrm{vec}(\mathbf{W}),
\]
and stacking all $N$ samples  into the vector $\mathbf{Q} = [\mathbf{q}_1^{\top}\;\;\cdots\;\;\mathbf{q}_N^{\top}]^{\top}$, we obtain the equivalent problem 
\begin{equation*}
\min_{\mathbf{W}}
\|
\mathbf{Q}
-
(\mathbf H \otimes \mathbf{I}_n)
\mathrm{vec}(\mathbf{W})
\|^2 ,
\label{eq:appendix_stacked}
\end{equation*}
where $\mathbf{H} = [\widehat{\boldsymbol{\eta}}_1\;\;\cdots\;\;\widehat{\boldsymbol{\eta}}_N]^{\top}$. Using the Kronecker product property $(A\otimes B)^\dagger=A^\dagger\otimes B^\dagger$, we obtain the least-squares solution 
\begin{equation}
\mathrm{vec}(\mathbf{W})
= (\mathbf H \otimes \mathbf{I}_n)^{\dagger} \mathbf{Q}=(\mathbf H^\dagger\otimes \mathbf{I}_n)\mathbf{Q},
\end{equation}
from which we directly also recover $\mathbf{W}_1$ and $\mathbf{W}_{2:M}$. Thus we only 
need to compute the pseudoinverse of the 
$N \times K$ matrix $\mathbf{H}$, which yields an efficient closed-form update at each iteration.

\phantom{.}
\newpage

\bibliographystyle{jfm}
\bibliography{ref}

\end{document}